\documentclass[preprint2]{aastex}

\newcommand{\msun}{$M_\odot$}

\newcommand{\rsun}{$R_\odot$}

\newcommand{\kms}{km~s$^{-1}$}
\newcommand{\mdotyr}{$M_\odot$~yr$^{-1}$}

\def \la{\mathrel{<\kern-1.0em\lower0.9ex\hbox{$\sim$}}}
\def \ga{\mathrel{>\kern-1.0em\lower0.9ex\hbox{$\sim$}}}

\shorttitle{The Binary Hypothesis for Planetary Nebulae }
\shortauthors{De Marco}

\begin{document}


\title{The Origin and Shaping of Planetary Nebulae:  \\ Putting the Binary Hypothesis to the Test}


\author{Orsola De Marco}
\affil{Astronomy Department,  American Museum of Natural History, New York, NY 10024}
\email{orsola@amnh.org}


\begin{abstract}
Planetary nebulae (PNe) are circumstellar gas ejected during an intense mass-losing phase in the the lives of asymptotic giant branch stars. PNe have a stunning variety of shapes, most of which are not spherically symmetric. The debate over what makes and shapes the circumstellar gas of these evolved, intermediate mass stars has raged for two decades. Today the community is reaching a consensus that single stars cannot {\it trivially} manufacture PNe and impart to them non spherical shapes and that a binary companion, possibly even a sub-stellar one, might be needed in a majority of cases. This {\it theoretical conjecture} has however not been tested observationally. In this review we discuss the problem both from the theoretical and observational standpoints, explaining the obstacles that stand in the way of a clean observational test and ways to ameliorate the situation. We also discuss indirect tests of this hypothesis and  its implications for stellar and galactic astrophysics.  
 \end{abstract}


\section{Introduction}
\label{sec:introduction}

In the current paradigm a planetary nebula (PN) is produced by a mass-losing asymptotic giant branch (AGB) star and later ionized by the AGB star core, on its way to becoming a white dwarf (WD). The great majority of PNe are not spherical. They display  axi- and point-symmetries, sometime with the addition of jet-like structures. These shapes have been attributed to the action of stellar rotation and/or magnetic fields \citep[e.g.,][]{GarciaSegura1999,GarciaSegura2005}, which result in an equatorially-concentrated AGB mass-loss. When the post-AGB star heats up, it blows a fast but tenuous wind, which is constrained at the equator, resulting in an elliptical or bipolar PN. In a minority of cases, the action of a companion is thought to be required in order to produce jets, point symmetries, or to explain off-center central stars (CSPN; see \citealt{Balick2002} for a review).

For the last two decades the community debated whether stellar rotation and global magnetic fields can be sustained in {\it single} AGB stars and thus confer non-spherical morphologies to PNe \citep[e.g.,][but see also \citealt{Bond1978}]{Soker1989,Soker1997,DeMarco2006,Zijlstra2007}. Recently, the debate has been rekindled \citep{Soker2006,Nordhaus2007} by the argument that in a large majority of cases, single AGB stars are unlikely be able to sustain large scale magnetic fields for long enough to affect shaping, because the field drains the star of angular momentum on short time scales and quenches itself. As the models of, e.g., Garcia-Segura and collaborators have shown, magnetic fields can be an effective ingredient in shaping many PNe, but in those models the magnetic field strength was assumed constant and was not coupled to the negative feedback action of the stellar envelope. Without the action of an angular momentum source, the field would vanish. Such an angular momentum source could effectively be provided by a stellar or sub-stellar companion. If this is what actually happens, 
then PNe would be, by and large, a binary-interaction phenomenon.

In addition to the problem of understanding how PN symmetries diverge from spherical, we also need to explain why they form in the first place. Stars approaching the final phases of their AGB life transit to a high mass-loss mode, effectively ejecting the remaining envelope in the space of a few hundred to a few thousand years. This heavy mass-loss, dubbed the ``super-wind", is observed in end-of-AGB stars and explains PN masses and their density structures. Although several ideas have been presented in the literature \citep[e.g.,][]{Lagadec2008}, we still do not know what triggers the super-wind. A companion interacting with the expanding AGB star could act as such trigger.

In this paper we review the problems of PN formation (triggering of the super-wind) and shaping (making the super-wind diverge from a spherical distribution) in the context of binarity. We dwell more on the shaping part of the problem, while leaving a review of the possible physical connections between binarity and super-wind trigger to future papers. We call the hypothesis that companions are the main cause of PN formation and shaping, the {\it Binary Hypothesis}.

{\it In the Binary Hypothesis, PNe form more readily around binaries, where by binary we intend a star accompanied by another star, a brown dwarf or even a planetary system. If this were the case, the CSPN population would derive primarily from those main sequence stars that have such companions, and would have a relatively higher multiplicity fraction or be often the product of mergers (although some PNe could still derive from single stars).}  

By contrast, we shall refer to the current model for the formation and shaping of PNe as the {\it Single Star Paradigm}. We stress that in the Single Star Paradigm, PN origin and shaping can be influenced by binarity, but binarity is not a favored channel and any other mechanisms are as likely to form and shape PN as binary interactions. 
Hence in the Single Star Paradigm, PNe are formed and shaped by single stars as well as binaries, but in different proportions compared with the Binary Hypothesis. 

Observations cannot yet discern which of the two paradigms is more correct. The obstacles to detecting binary CSPNe, in particular for periods longer than a few weeks, are substantial. As a result, the CSPN binary fraction is not well known. Planets and brown dwarfs have never been detected around CSPNe and it is unlikely that many will be detected in the near future around these generally distant targets. In addition, the overall number of confirmed binary CSPNe that could be used to relate binarity and PN morphology is rather small.  Finally, it is already clear that many nebulae we call today PNe are more related to other phenomena. These PN mimics \citep[e.g.,][]{Miszalski2009,Frew2008b} can complicate the task of an observational test PN origin. Yet, this issue needs to be tackled observationally\footnote{To tackle this problem a collaboration, dubbed PlaN-\"B for Planetary Nebula Binaries, was forged at the Asymmetric PN IV meeting, that took place in La Palma, Spain, in June 2007 -- www.wiyn.org/planb/.}: we need to obtain a statistical sample of binary CSPNe free from PN mimics and study these binaries in relation to their PNe. We also need to develop techniques to detect substellar companions and to identify past mergers indirectly.




In this paper we review the PN Binary Hypothesis, starting with its theoretical basis (\S~\ref{sec:theoreticalconsiderations}). We then review (\S~\ref{sec:theknownbinaries}) what is currently known about CSPNe in binary systems. In \S~\ref{sec:predictionsandtests} we review direct and indirect predictions and tests of the PN Binary Hypothesis. In \S~\ref{sec:progenitorsandprogeny} we cast the PN population in the context of its progenitors (the AGB and post-AGB stars) as well as progeny (the WDs),  highlighting how the binary characteristics of these two populations enlighten and complicate the Binary Hypothesis for PNe.  In \S~\ref{sec:relatedclasses} we bring related binary classes, such as cataclysmic variables, to bear on the Binary Hypothesis. We then (\S~\ref{sec:additionalriddles}) report a few additional challenges facing any paradigm for the formation of PNe. We conclude in \S~\ref{sec:conclusions}. Throughout this review, we have made extensive use of Bruce Balick's Osterbrock PN Image Catalogue\footnote{http://www.astro.washington.edu/balick/PNIC/}, to which we refer the reader for images of the many PNe discussed.


\section{Considerations on the \\ need for a companion}
\label{sec:theoreticalconsiderations}

In this Section we will review the debate over what are the actual causes that shape PNe into axi-symmetric, point-symmetric and asymmetric geometries and why a companion might be called for (see also \citealt{Balick2002} and \citealt{Soker1997}).

\subsection{From AGB to PN: the super-wind and the mass-loss geometry}

Stars between $\sim$1 and 10~\msun\ undergo two phases of expansion, one after core hydrogen runs out, called the red giant branch (RGB) and one following core helium burning, called the AGB
\citep[e.g.,][]{Vassiliadis1994,Herwig2000,Iben1995}. 

At the end of the AGB phase the stellar wind  (speeds of 10-15~\kms, mass-loss rates $\sim$10$^{-7}$~\msun~yr$^{-1}$) unexpectedly increases in intensity with mass-loss rates surging to as much as 10$^{-4}$~\msun~yr$^{-1}$ \citep[][]{Delfosse1997}. This super-wind phase quickly depletes the AGB envelope till, when the envelope mass falls below 10$^{-3}$--10$^{-4}$~\msun, the star structure changes, the photospheric radius shrinks and the stellar effective temperature rises. The mass-loss rate of the star drops at this point to $\sim$10$^{-8}$~\mdotyr\ and the wind speed increases to $\sim$200-2000~\kms\ \citep{Perinotto1989}. The fast wind plows up the circumstellar material ejected during the super-wind phase and the resulting circumstellar gas distribution is then ionized by the heating central star. The super-wind is assumed (and observed, e.g., \citealt{Castro-Carrizo2007}) to depart from spherical symmetry so that the resulting nebula is not spherical. This model is known as the Interactive Stellar Wind (ISW) model \citep{Kwok1978}. It and its later modifications, known under the name of Generalized ISW \citep[GISW;][]{Icke1988,Icke1989,Icke1992,Soker1989,Frank1994,Mellema1994,Mellema1995a,Mellema1995b,GarciaSegura1999} explain well PN gas density structure, kinematics and morphology (at least for the main body of the PNe). However, the causes of the super-wind triggering, and geometry are {\it assumed}, not explained.



\begin{figure*}
\includegraphics{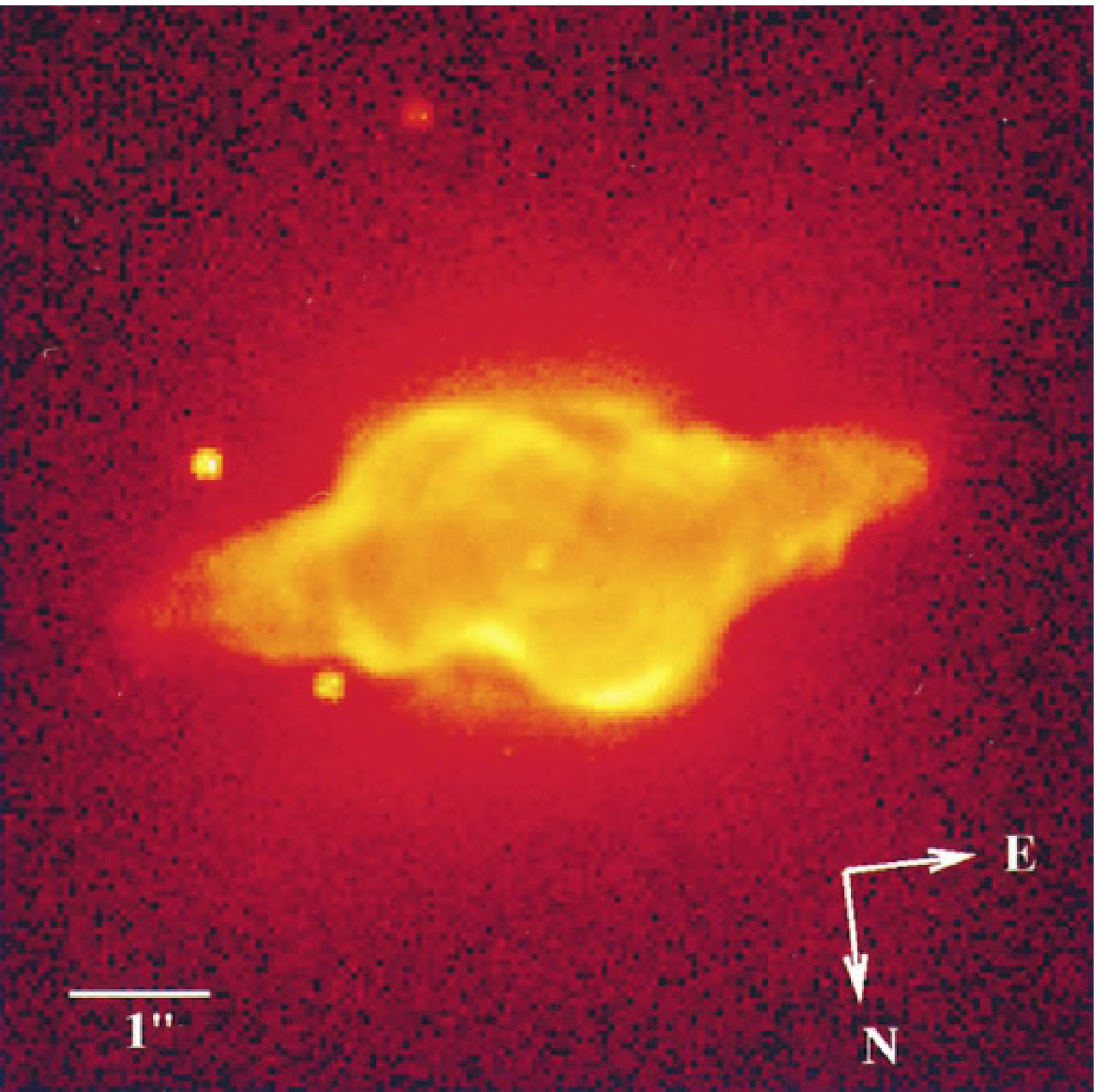}
\includegraphics{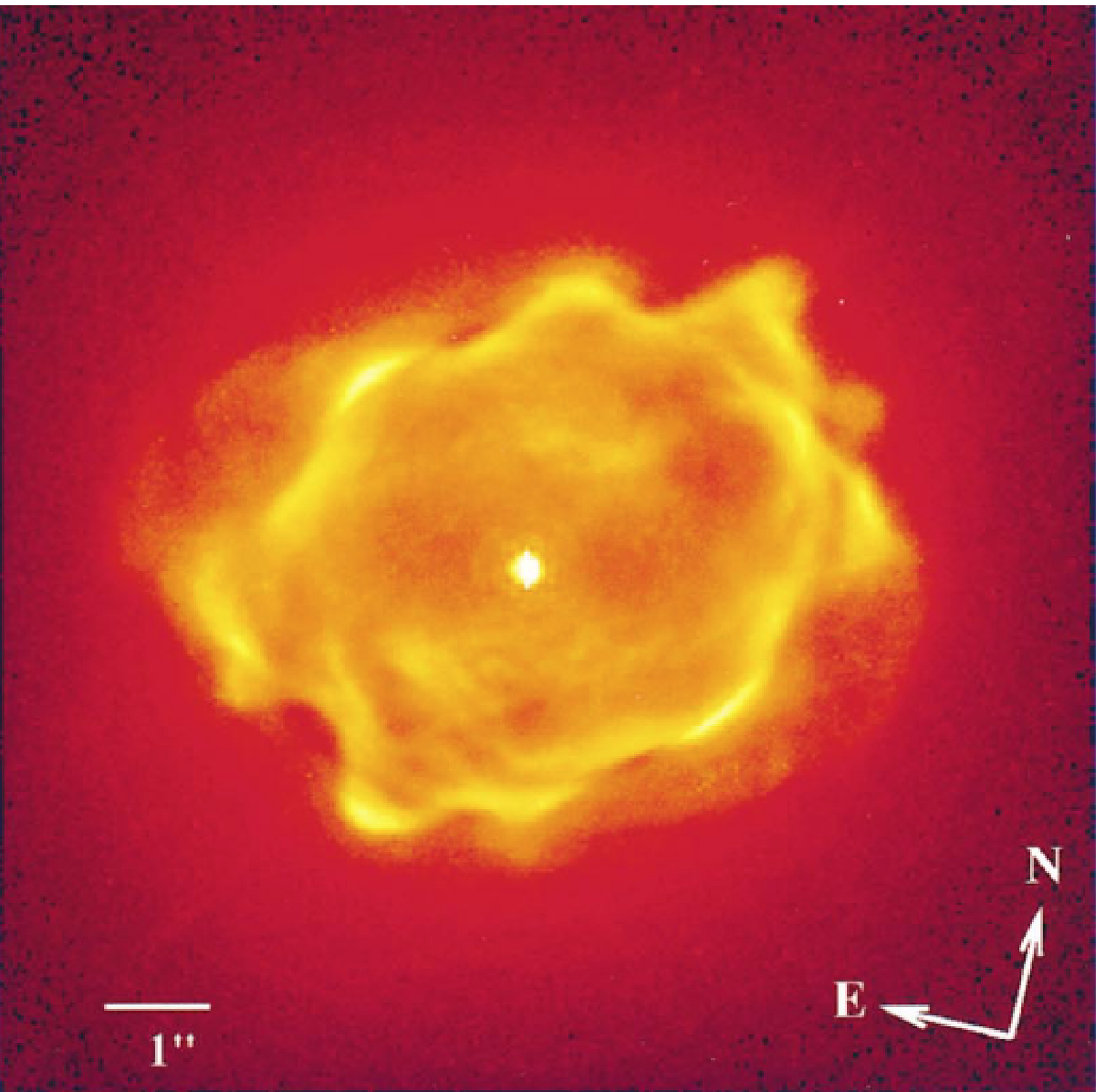}
\includegraphics{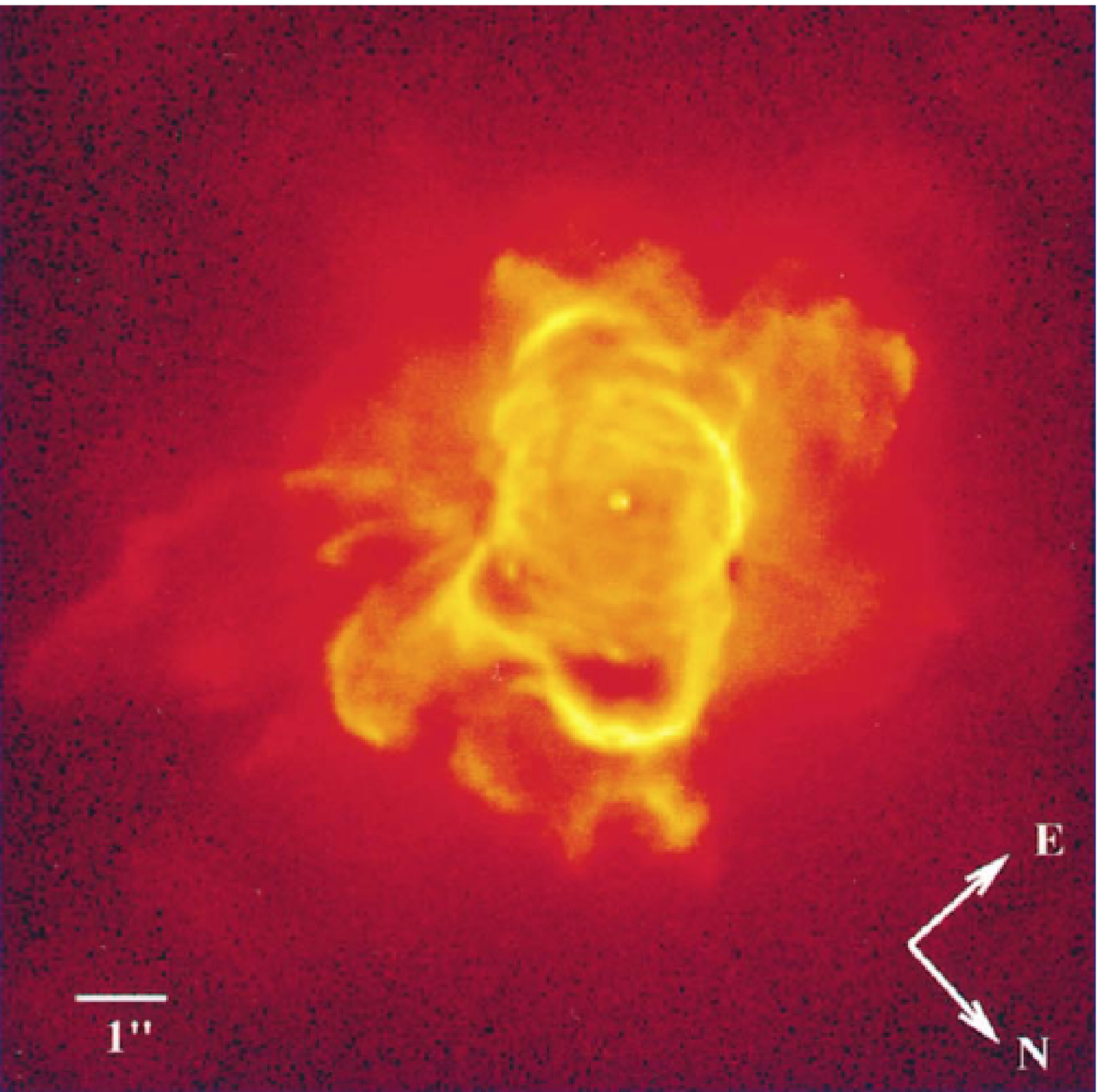}
\caption{HST H$\alpha$ images of very young PNe (He~2-115, left, He~2-138, center and M~1-26, right) demonstrating the extreme morphologies exhibited by these objests. From \citet{Sahai1998}, reproduced by permission of the AAS.}
\label{fig:youngpne}
\end{figure*}

A complication was added when we learned that shapes of very young PNe are so complex that even assuming an equatorially concentrated AGB mass-loss would not be enough to explain them. \citet{Sahai1998} demonstrated that only if the {\it post-AGB} mass-loss is heavily collimated  these shapes could be explained (see Fig.~\ref{fig:youngpne}). Once again, these collimated outflows were {\it assumed}.
Finally, \citet{Bujarrabal2001} determined that most pre-PNe (nebulae that are not yet photo-ionized and shine by starlight reflection or shock-ionization) have outflow linear momenta much in excess of what could be driven by radiation pressure from single stars.
  
We can all agree, that a complete model for the evolution and shaping of PNe, needs to {\it explain} the surge in mass-loss rate called the super-wind, and the origin of AGB and post-AGB mass-loss geometries, rather than {\it assume} them.  



\subsection{``Derived" shaping agents: 
Stellar rotation and magnetic fields}
\label{ssec:secondaryshapingagents}

Non-spherical AGB mass-loss geometries have been traditionally ascribed to ``stellar rotation, magnetic fields and binarity". It has however been pointed out by \citet{Soker1997} that not all these can be the original causes. For instance, if the AGB magnetic field is caused by a  companion that spins up the envelope, we would call the binary the actual, or {\it original cause} and the magnetic field the {\it derived cause} of the shaping.


\citet[][see also \citealt{GarciaSegura2005} and \citealt{GarciaSegura2006}]{GarciaSegura1999} demonstrated that if an AGB star is assumed to be rotating and/or to possess a large-scale magnetic field, it can eject an equatorially enhanced super-wind, which can then lead to axi-symmetric PNe. 
\citet{Soker2006} and \citet{Nordhaus2007}, however, showed that the differential rotation needed to produce such a field, is drained by the field itself. As a result the magnetic field is too short-lived ($\sim$100~years) to affect the AGB super-wind geometry. The models of \citet{GarciaSegura1999} contained no feedback of the AGB envelope gas onto the magnetic field strength, so the field's intensity was  assumed constant. If indeed a single AGB star cannot sustain a global magnetic field (but see also Sec.~\ref{ssec:primaryshapingagents}), then a single AGB star cannot shape an axi-symmetric PN with it. 

\citet{GarciaSegura1999} argue that stellar rotation alone can shape the super-wind and that the rotation rates needed to impart an axial geometry are actually quite small if the super-wind is triggered by an instability of the pulsating AGB star. Although, the smallest viable value of the rotation rates cannot be predicted without a quantitative model of the instability that causes the super-wind, they are of the order of a few \kms\ and \citet{GarciaSegura1999} argue that such rates can be achieved by stars more massive than 1.3~\msun. Observational values of AGB rotation rates point to upper limits of a few km~s$^{-1}$  (e.g., \citealt{deMedeiros1999}\footnote{The exact number is relevant but hard to establish without knowing the exact evolutionary stage of the spectral classes included in the observations. In addition, we really want to know the rotation rate of stars in the super-wind phase, which are very dusty and optically  very faint.}), therefore potentially sufficient. However, \citet{Ignace1996}, based on different physical reasoning, declared it unlikely that rotation rates in single AGB stars would be high enough to contribute to a nonspherical geometry. We are clearly far from a resolution. Finally, one wonders whether the assumption of rotation alone in a model can be considered physical at all; a single rotating and convecting giant can hardly lack a magnetic field, which would then slow the rotation down by the argument above. 


\subsection{``Original" Shaping agents: binary interactions and magnetic spots}
\label{ssec:primaryshapingagents}

Binaries close enough to interact during the AGB phase of the primary, have naturally multiple ways in which they can shape the AGB and/or the post-AGB mass-loss both directly, by the action of gravity, or indirectly, by stimulating secondary phenomena such as pulsations or magnetic fields. Following \citet{Soker1997}, we distinguish five main types of PN-shaping binary interactions and match them to PN morphologies: 

{\it (i) Very wide binaries,} where the orbital period is much longer than the lifetime of the PN, could at best produce some bubbles or other small features in the PN (e.g., NGC~246).

{\it (ii) Wide binaries,} where the length of the orbital period is of the same order of the lifetime of the PN ($\sim$100 to $\sim$1000~AU), are the most likely to produce deviations from axi-symmetry, such as bent jets (e.g., NGC~6826).

{\it (iii) Closer binaries that avoid a common envelope} (separations in the few to $\sim$100~AU range; the lower limit is very uncertain) can result in a variety of PN shapes, depending on parameters such as separation and mass-ratio. For instance, spiral structures can be driven into the AGB mass-loss \citep{Edgar2008}. Elliptical orbits could stimulate episodic mass-loss during apastron, and result in a CSPN that is not in the middle of the PN (e.g., MyCn~18). Within this class, \citet{Manchado1996b} suggest that the interactions of disks and companions can cause precessing jets, which can inflate more than a pair of lobes resulting in quadrupolar PNe. Bipolar PNe can be caused by wind accretion (this is in line with evidence from symbiotic nebulae -- see \ref{ssec:symbiotics}). Elliptical PNe can result if the separation is larger and/or companion mass lower. A companion might avoid a common envelope, and just interact with the envelope of the primary during the lower AGB, but enter a common envelope at a later time, due to further expansion of the AGB star. This two-stage interaction is likely to result in a more complex morphology.

{\it (iv) Common envelope interactions where the binary survives.} The common envelope interaction, happens when an RGB or AGB star transfers mass onto a companion at a rate that is too large to be accreted. The companion then expands, fills its own Roche Lobe and the two stars quickly become engulfed by the primary's envelope. The secondary transfers energy and angular momentum to the primary and can unbind the envelope \citep{Paczynski1976,Iben1993}. If the companion is able to eject the envelope the result is a close binary. Although low mass companions have a higher chance of merging with the core of the primary, there is evidence that even brown dwarfs can eject giants' envelopes \citep{Maxted2006}. The common envelope interaction is extremely complex and it is at this point difficult to predict the circumstellar environment of such post-interaction binaries, except maybe that the mass is lost preferentially on the equator \citep{Sandquist1998,DeMarco2003}, resulting on an elliptical or bipolar PN. \citet{Soker1997} conjectures that the resulting PN will be elliptical instead of bipolar, partly on the argument that most PNe around known post-common envelope binaries are not bipolar. However, we will argue in Sec.~\ref{ssec:morphologykinematics} that many of the morphologies are actually bipolar, where the lobes have faded, or where the PN is seen pole-on.


{\it (v) Common envelope interactions that result in a merger.} If the common envelope interaction with a companion does not result in the ejection of the envelope, the companion merges with the core of the primary. This can be the case for both stellar and sub-stellar companions, depending on the binding energy of the primary's envelope and the mass of the companion. If the companion becomes tidally shredded as it approaches the core of the primary it could form a disk around it \citep{Nordhaus2006}. Such a disk can cause the ejection of jets, while the overall PN morphology could be elliptical due to spinning up of the envelope.

Both stellar and substellar companions can excite pulsational modes when they enter a common envelope phase with the primary \citep[and, in the case of sub-stellar companions, they are eventually destroyed;][]{Soker1992}.  In both these common envelope cases, \citet{Nordhaus2007} showed that a strong magnetic field will also result \citep[see also][]{Tout2008} that can contribute to shaping.

Reliable numerical models of binary interactions are still at the cutting edge of computing technology \citep[for a modern effort consult, e.g.,][]{Ricker2008}. Models that simulate the full binary interaction as well as the PN ejection and resulting morphology do not exist. Numerical models of PN formation from a close binary that avoids a common envelope phase are presented by \citet{Garcia-Arredondo2004}, where they assumed that jets are blown by a companion that is accreting wind from a primary at 10~AU,  and by \citet{Gawryszczak2002}, who considered mostly gravitational focussing of the AGB mass-loss.

In a scenario not involving binaries, \citet{Soker1999} proposed that magnetic spots on the surface of a single AGB star can be the {\it original} cause leading to elliptical PN shapes, though not strongly bipolar ones.  

Finally, we cannot exclude that some single AGB 
stars could sustain a global magnetic field unaided. The Sun sustains its magnetic field by tapping convection energy to resupply differential rotation \citep{Rudiger1989}.  Such steady dynamo operating during the AGB phase was investigated as an {\it original cause} of bipolar PN shapes \citep{Blackman2001}.  However, the dynamo in that study was weak, so that the magnetic field had to last until the end of the AGB phase, somewhat unrealistically, since by the end of the AGB phase the stellar envelope is nearly depleted.  In a continuation of this work, \citet{Nordhaus2007} incorporated the back-reaction of the magnetic field growth on the stellar rotation profile in order to investigate a more physical AGB dynamo.  In particular, they showed that convective reseeding (and hence a sustained dynamo) could be achieved if a couple of percent of the convective energy could be tapped in analogy to the solar model.  In their work, the field needed to be stored in the shear layer, with minimal buoyant rise, until the aggregate Poynting flux was large enough to unbind the AGB envelope.  As they discussed, it remains to be seen how viable such convective resupply model can be in AGB stars.

However, the {\it raison d'\^etre} of this review is not whether binaries {\it can} shape PNe more easily than single stars, but whether {\it they do}! Hence, from here on we will concentrate on observational tests of the PN Binary Hypothesis.

  
\section{Current knowledge of binarity in PNe}
\label{sec:theknownbinaries}

In this Section we will describe what is known about binarity in PNe from surveys. We will also describe the characteristics of the handful of confirmed binary CSPNe and their PNe. Finally, we will outline which PN and CSPN characteristics make some objects very compelling binary candidates. A sizable binary CSPN sample is fundamental to relate binarity and shaping as well as to design better binary-finding methods. Today, the least massive companion to a CSPN has a late M spectral type, so we have no test of the influence of sub-stellar companion on PN origin and shaping. 

\subsection{Binary central star searches}
\label{ssec:binarycentralstarsearches}

\citet{DeMarco2008c} have reviewed the efforts of Howard Bond and collaborators \citep[e.g.,][]{Bond1979,Bond1992,Bond1995,Bond2000} who used a photometric variability technique to  detect binaries via periodic light variability due to irradiation of a cool companion by the hot primary, ellipsoidal variability or eclipses. This survey has taken place over three decades and determined that 10-15\% of about 100 monitored CSPNe are close binaries with periods shorter than 3 days (though most systems have periods smaller than 1 day). This survey technique detects more readily binaries with short periods, since, as the orbital separation increases, the size of the irradiation effect, the ellipsoidal variability and the likelihood of eclipses all diminish. \citet{DeMarco2008c} calculated irradiation models and used them to argue that, for average system parameters, the longest period for binaries discovered by a survey such as that of Bond and collaborators, should be of the order of 2 weeks (for a detection limit of $\sim$0.1~mag), which is longer than the longest period detected by that survey (3 days). They therefore proposed that either the survey suffered an additional bias that prevented the detection of longer period binaries, or, that post-common envelope binaries really have only very short periods ($<$3~days), i.e., binaries with slightly longer periods (e.g., 3~days $<$ P $<$ 2 weeks) are very rare. 

\citet{Miszalski2008b} and \citet{Miszalski2009} used the {\it Optical Gravitational Lensing Experiment (OGLE) II and III} survey to carry out a similar search for periodic photometric variables. They detected 21 periodic variable CSPNe, where the variability could be ascribed to irradiation effects, ellipsoidal variability or eclipses. After 20 years of slow progress they doubled the CSPN close binary sample. Compared to the survey of Bond and collaborators the technique of \citet{Miszalski2009} insured a much higher level of homogeneity. With a variability detection limit similar to that of Bond and collaborators, \citet{Miszalski2009} determined a similar close binary fraction (12-21\%) and period distribution (almost no systems have periods larger than 3~days). Using the entire close binary CSPN sample, \citet{Miszalski2009} confirmed that the dearth of binaries with periods longer than $\sim$3~days is not due to a bias but is instead a feature of the common envelope interaction, as suspicion already voiced by \citet{DeMarco2008c} based on irradiation models and half the number of confirmed binaries.

\citet{Miszalski2009} noticed that the only population synthesis prediction that fits the data is that of \citet{deKool1992}. All other calculations predict many more systems at longer periods. The only difference between the calculation of \citet{deKool1992} and the others \citep[e.g.,][see also \S~\ref{ssec:wds}]{Yungelson1993,Han1995,Politano2007} appears to be the choice of a binary mass ratio distribution; \citet{deKool1992} choses both star's masses independently using the initial mass function. The other calculations use a flat mass ratio distribution. Neither of these choices conforms with current knowledge \citep[which finds a mass ratio distribution dependent on $(M_2/M_1)^{-0.5}$;][]{Duquennoy1991,Shatsky2002}, so it will be interesting to see how future predictions fit the observed binary central star period distribution.

Another method to search for binaries is via periodic radial velocity (RV) variability of the stellar lines. Over the last 20 years a few surveys have used this method (Table~\ref{tab:RVsurveys}). \cite{Mendez1989} took one or two high resolution spectrograms of each CSPN in a sample of 28 objects and concluded that none was variable beyond doubt, in particular since several objects might be wind variables. \cite{Sorensen2004} carried out a survey at intermediate resolution, and determined that 39\% of their sample were RV variables. Finally, \citet{DeMarco2004} and \citet{Afsar2005} detected 91\% and 37-50\% RV variable fractions, respectively (although their sample sizes were undoubtedly small). The smaller RV variable fraction of the \cite{Afsar2005} survey is due to the lower resolution of their setup ($\sim$10~\kms\ instead of $\sim$3~\kms), which is also similar to the resolution used by \cite{Sorensen2004}.  The real caveat of {\it all} these  surveys is that no periods were detected so that wind variability is still an alternative explanation for the presence of RV variability in several objects.

\begin{figure*}
\includegraphics{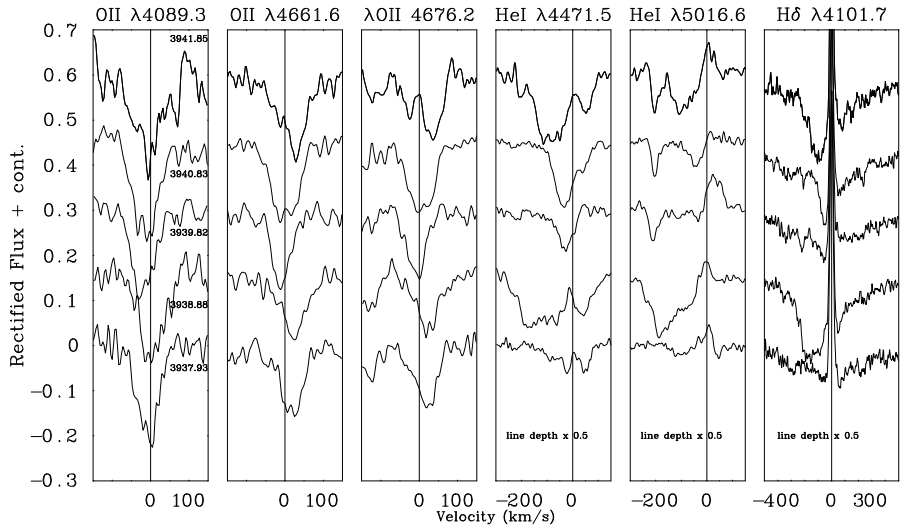}
\caption{Helicentrically-corrected spectra of the CSPN of M~1-77 taken on 5 subsequent nights (on the left panel we report the dates as Julian Date - 2\,450\,000 days). The emission component to H$\delta$ (right panel) derives from the PN. Adapted from \citet{DeMarco2008a}.}
\label{fig:m1-77}
\end{figure*}

In an attempt to detect periods as well as assess the impact of intrinsic wind variability \citet{DeMarco2008a} obtained echelle-resolution spectrograms of four of the 10 RV-variabile CSPNe of \citet{DeMarco2004}. After one observing run, the RV variability of the stellar spectral lines of two objects appeared to exhibit tantalizing sinusoidal behavior \citep{DeMarco2006}. After a second observing run the behavior could not be confirmed, so that \citet{DeMarco2008a} tentatively ascribed it to wind as well as pulsation-induced variability \citep[e.g.,][]{Patriarchi1997}.   
However, none of the RV variability observed in the four objects clearly matched these supposed causes (e.g., emission lines in the CSPN of IC~4593 were found to shift towards the red as well as the blue, contrary to normal wind-induced behavior; the heliocentric RVs of the spectrum of BD+33~2642 in 2005, had a positive average wavelength, instead of zero, as expected for binarity, but in 2006 the average had moved to a negative wavelength! Finally, the extreme RV variability of the absorption lines of the {\it non-windy} CSPNe of M~2-54 and M~1-77 [e.g., Fig.~\ref{fig:m1-77}] could not readily be explained by pulsations).  

It is therefore possible that the observed spectral variability, though not straight-forwardly related to binarity, is {\it indirectly} caused by the presence of a companion. A companion can interact with the wind of the primary by accreting it or just by moving through it thus causing an accumulation of material on the bow shock formed around the companion as is sometime the case in massive binaries \citep[][]{DeMarco2002c}. \citet{Izumiura2008} inferred the presence of a low mass main sequence or WD companion to the silicate carbon AGB star BM Gem. Their suggestion hinges on the presence of a substantial blue continuum and Balmer lines with variable PCygni profiles. They conclude that the companion is accreting AGB wind material and blowing an outflow. While the situation for CSPNe is undoubtedly different, in that their radii are much smaller, and mass-loss rates lower, it is possible that a parallel can be found. If so, we might find a way to detect companions indirectly.

Efforts are continuing to find binaries via RV variability work. De Marco et al. (in preparation) have used the Very Large Telescope to target faint CSPNe, where winds and pulsations should not be important. Negative detection of even a small sample with these characteristics should be able to impose some stringent constraints.  

Finally, it is possible to detect cool companions by infrared (IR) excess surveys \citep[][]{Frew2008}. This technique would however struggle to detect mid-to-late M-type companions since CSPNe, even if hot, are still very luminous in the near IR, where they can easily dominate faint M stars (but see Sec.~\ref{sssec:infraredexcessbinaries} and \ref{ssec:fracanddist}).

\subsection{Binary central stars: stellar and system characteristics}
\label{ssec:stellarandsystemcharacteristics}

In Tables~\ref{tab:knownbinaries} and \ref{tab:widebinaries} we list all the known binary CSPNe. We consider a confirmed binary only the following categories: periodic photometric variability ascribable to irradiation of a cool companion by the hot CSPN, ellipsoidal variability due to either or both of the stellar components filling, or partly filling their Roche lobes, or eclipses; periodic radial velocity variability, composite spectra or colors well understood as two separate stars and, finally, wide binaries with a low probability of a chance alignment. We note that we have included a few cases where both stars in a binary CSPN are too cool to be the ionizing source of the PN (e.g., SuWt~2 and, possibly, M~3-16 and M~2-19). In such cases it is suspected that the hot CSPN is an undetected tertiary component in the system.  

Several CSPNe detected because of photometric variability have been discussed in detail by \citet{DeMarco2008c}, where references for the values cited in Table~\ref{tab:knownbinaries} can be found. 
Hb~12 has been removed from the binary list of \citet{DeMarco2008c}, because its photometric variability does not appear as regular as previously announced (Hillwig, priv. comm). This system might be eclipsed by orbiting dust as is the case for the binary CSPN of NGC~2346, and it therefore remains a strongly suspected binary (see Sec.~\ref{ssec:suspectedbinaries}). In addition, after the publication of the paper by \citet{DeMarco2008c}, A~41 was confirmed to be an ellipsoidal variable by \citet{Shimanskii2008} and not an irradiated system.  Finally, 21 new photometric binaries have been identified by \citet[]{Miszalski2009} (we have excluded one of the binaries found in their list: the CSPN of M~1-34 is not a periodic variable and does not conform to our binary criterion). For the 21 objects discovered by \citet{Miszalski2009}, we do not report any parameters, though some do exist in the literature, to emphasize that consistent binary models for these objects have not yet been calculated. Below we discuss composite-spectrum objects, near IR excess and visual binaries.

\subsubsection{Composite spectrum binary central stars}
\label{sssec:compositespectrumbinaries}

Four CSPNe have composite spectra that point to the presence of a hot star and a cool companion. All four objects have very large low surface brightness PNe, which one might classify as elliptical but which also suffered interaction with the interstellar medium (ISM). Based on their PN morphologies, \citet{Soker1997} classified them as being due to a common envelope with a stellar companion. Before listing them, we warn that the presence of an {\it evolved} companion to the hot CSPN in at least three of the PNe below is suspicious, making these objects more akin to symbiotic systems (see Sec.~\ref{ssec:symbiotics} for an elaboration of this concept).

{\it A~35.} The spectrum of the CSPN of A~35 is composed of a hot star \citep[T=(80\,000$\pm$3000)~K;][]{Herald2002} and a cool evolved component with spectral type G8~III-IV. \citet{Gatti1998} resolved these two components with the Hubble Space Telescope (HST) and determined a separation of (18$\pm$5)~AU for a distance of 160~pc (we could have listed this CSPN also in Table~\ref{tab:widebinaries}). This is in line with the lack of radial velocity variability \citep{Gatti1997}. The cool star in A~35, like the one in LoTr~5, (see below) has a very high $v \sin i$ = 90~km~s$^{-1}$ \citep{Vilhu1991} and is thought to have accreted angular momentum from the wind of the AGB progenitor of the primary \citep{Jeffries1996}. Wind accretion is also the reason likely to explain the fast-rotating main sequence stars called FK Comae \citep{Walter1982}. \citet{Frew2008b} exclude that the PN and central binary are associated, implying that the nebula is a Str\"omgren sphere ionized by the hot star in the binary. The hot star in the system appears however to have been a CSPN in the near past.

{\it LoTr~5.} The central star of this PN was discovered to have a composite spectrum consisting of a very hot component \citep[150\,000~K;][]{Feibelman1983} and a cool and evolved, chromospherically active G5~III companion. It has also been suggested, but never confirmed, that the system might be a triple with a third stellar component with spectral type M5 \citep{Malasan1991}. As is the case for A~35, the cool star rotates rapidly ($v sin i \sim 60$~\kms).

{\it LoTr~1.} Not much is known about this object, except that the K-type optical spectrum is that of a chromospherically active star (Mg II and Ca II lines in strong emission) and that the IUE UV spectrum is that of an extremely hot star, presumed to be a close binary companion to the cooler object \citep{Bond1989}.

{\it NGC~1514.} This PN contains a hot star (T$>$60\,000~K) and a cool, evolved companion of spectral type A0-3~III. The separation of the two stars is smaller than 0.1''. RV variability was reported in the literature, but could not be confirmed by subsequent studies \citep{Feibelman1997}. The UV continuum has brightened by a factor of two between 1978 and 1989, a fact that could be explained by dust obscuration and that would make the central binary similar to the confirmed binary CSPN of NGC~2346.

\subsubsection{Infrared excess binary central stars}
\label{sssec:infraredexcessbinaries}

If a hot CSPN has a near IR flux in excess of what can be predicted based on its spectral class (as determined by a full optical spectral analysis, or, at least an analysis of several blue optical colors) we can suspect that a cool unresolved binary companion is near it. A caveat of this search technique is hat  hot dust, or a non standard stellar atmosphere (e.g., a hydrogen-deficient one) can produce a near-IR excess, so that spectra or colors in multiple IR bands are needed.

\citet{Frew2008} have detected a near IR excess in several CSPNe, using the {\it Deep Near Infrared Survey (DENIS)} and the {\it 2-$\mu$m All Sky Survey (2MASS)} and used it to obtain statistical information on the presence of companions (see Sec.~\ref{ssec:fracanddist}), although their (unpublished) objects need to be confirmed. \citet{Zuckerman1991} detected two CSPNe to have IR excess. The CSPN of A~63 (which was at the time already known to be an eclipsing binary) and the CSPN of EGB~6, which was later resolved with the HST and that we list with the visual binaries in Table~\ref{tab:widebinaries}. 

\begin{figure}
\includegraphics{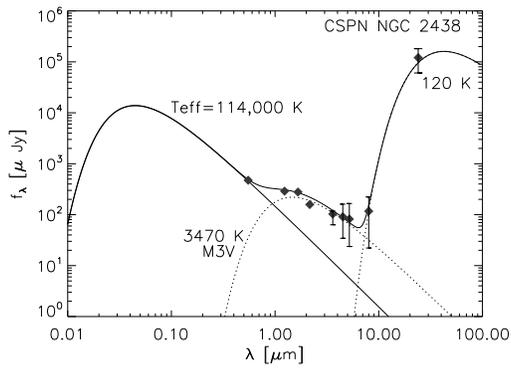}
\caption{The red and IR fluxes of the CSPN of NGC~2438 fitted with three blackbody curves and implying the presence of an M3V companion. Adapted from \citet{Bilikova2008}.}
\label{fig:ngc2438}
\end{figure}

Another CSPN discovered to be a binary via IR excess is in the middle of PN NGC~2438. We list it in Table~\ref{tab:knownbinaries} because its IR excess was measured over multiple bands ($J$, $H$, $K$ and the Spitzer/IRAC bands at 3.6, 4.5, 5.8, 8.0~$\mu$m; \citealt{Bilikova2008}), and can be fitted well by a blackbody curve with a temperature of 3470~K, making the companion an M3V star (Fig.~\ref{fig:ngc2438}). A spectroscopic analysis of the hot component was carried out by \citet{Rauch1999}, whose spectra only extended to $\sim$5500~\AA\ and could not have detected the cool companion.

\subsubsection{Visual binary central stars}
\label{sssec:widebinaries}

In Table~\ref{tab:widebinaries} we have listed visual binary CSPNe from the literature. Many were discovered by \citet{Ciardullo1999}'s HST survey. They labeled them as either ``probable" or ``possible". They found that between 9\% (``probable") and 14\% (including ``possible" associations) of all surveyed systems have a visual companion. Their projected separations (excluding limits) range from 160 to 2400 AU (for the ``probable" associations, but companions are found as far as 10\,580 AU from the primary for systems they call ``possible" associations). This survey is biased to separations larger than the HST resolution of $\sim$0.05''-0.1" and small enough that confusion with background or foreground objects does not excessively reduce the probability of association. Additionally, there is a bias against fainter companions (as can be seen in Table~\ref{tab:widebinaries}, all companions but one have spectral type K or earlier).  

The morphologies of the PNe around visual binaries are variable with at least one quite circular PN (A33; \citealt{Manchado1996}). However there does appear to be a common trend with several PNe exhibiting an inner ring (or bubble) and an outer more diffuse elliptical structure, which in a few cases tends towards bipolar \citep[e.g., Mz~2,][]{Gorny1999}. 

Based on their morphologies, \citet{Soker1997} lists most of these PNe as objects that suffered a common envelope with a substellar companion that was around the CSPN progenitor. Another possibility is that these visual binary CSPNe are today triple systems, where the undetected member is much closer to the primary and has interacted with it (A~63 is in fact a triple system, cf. Tables~\ref{tab:knownbinaries} and \ref{tab:widebinaries}).  

Leaving aside the hard-to-test suggestion that a sub-stellar companion was present, we ponder whether most visual binary CSPNe evolve from main sequence stars in similarly wide binaries, or whether they derive from the main sequence triples. If the progenitors of the visual binary CSPNe are the wide main sequence binaries, the main sequence period distribution of \citet{Duquennoy1991} would lead us to expect a larger number of visual binary CSPNe at or near the HST resolution limit than was found by \citet{Ciardullo1999}. The caveat in this statement is the small number of known visual binary CSPNe, which hinders a comparison between their projected separations and the periods of the main sequence binaries - one has to transform separations into periods with a set of statistical assumptions which are more reliable for larger samples.

In conclusion, finding  additional visual binary CSPNe will enable us to determine whether their progenitors are the main sequence binaries or the main sequence triples and multiples. In the first instance, the finding would support that non-spherical PNe can originate from single star mechanisms or that substellar companions are at play, while in the latter case the Binary Hypothesis would obtain some support.

\subsection{PNe around binary central stars: morphologies, kinematics and abundances}
\label{ssec:morphologykinematics}

{\it Morphologies and kinematics.} \citet{Bond1990} and \citet{Zijlstra2007} addressed specifically the morphologies of PNe with close binary CSPNe, agreeing that they have peculiar, uncommon features.  \citet{Bond1990} remarked that these PNe do not show the common multiple structures expected to derive from subsequent evolutionary phases and concluded that this is due the fact that the AGB evolution was interrupted by the common envelope interaction. \citet{Zijlstra2007} remarked that one might expect to find more bipolar PNe around this type of close binary central star, but this is not the case. \citet{Soker1997} instead expects that bipolar PNe should be produced by those binaries that avoid the common envelope phase, rather than by binaries that enter a common envelope interaction, which are the type of binaries usually detected inside PNe.

\begin{figure*}
\includegraphics{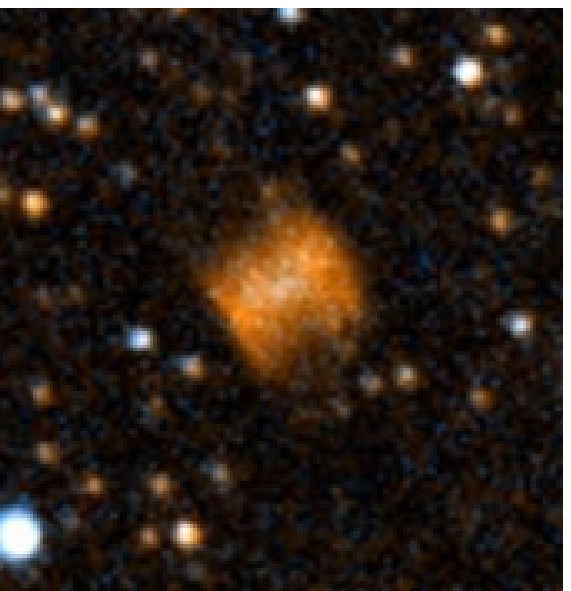}
\includegraphics{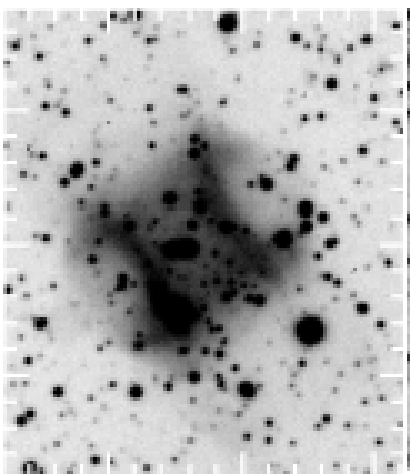}
\includegraphics{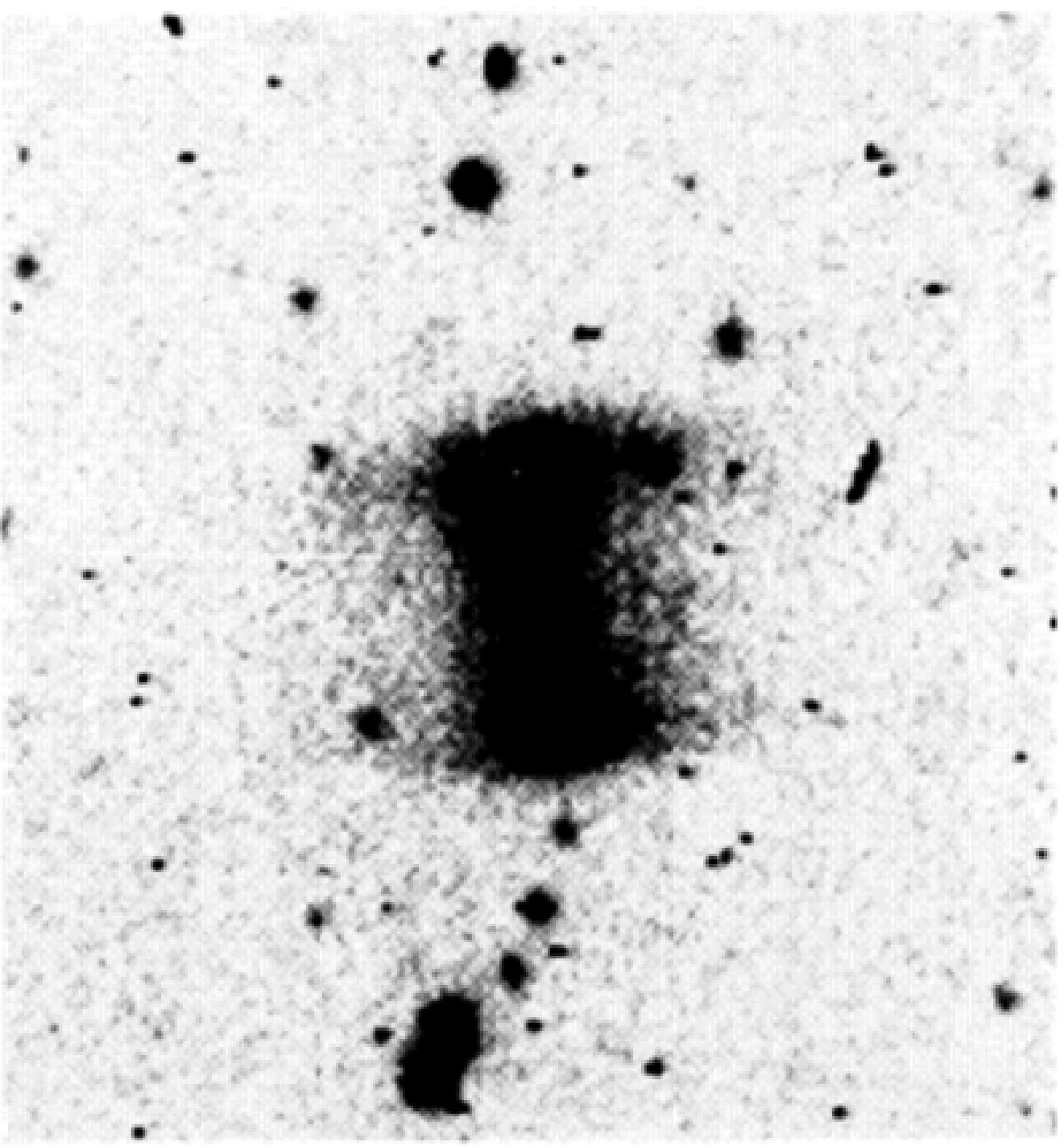}
\includegraphics{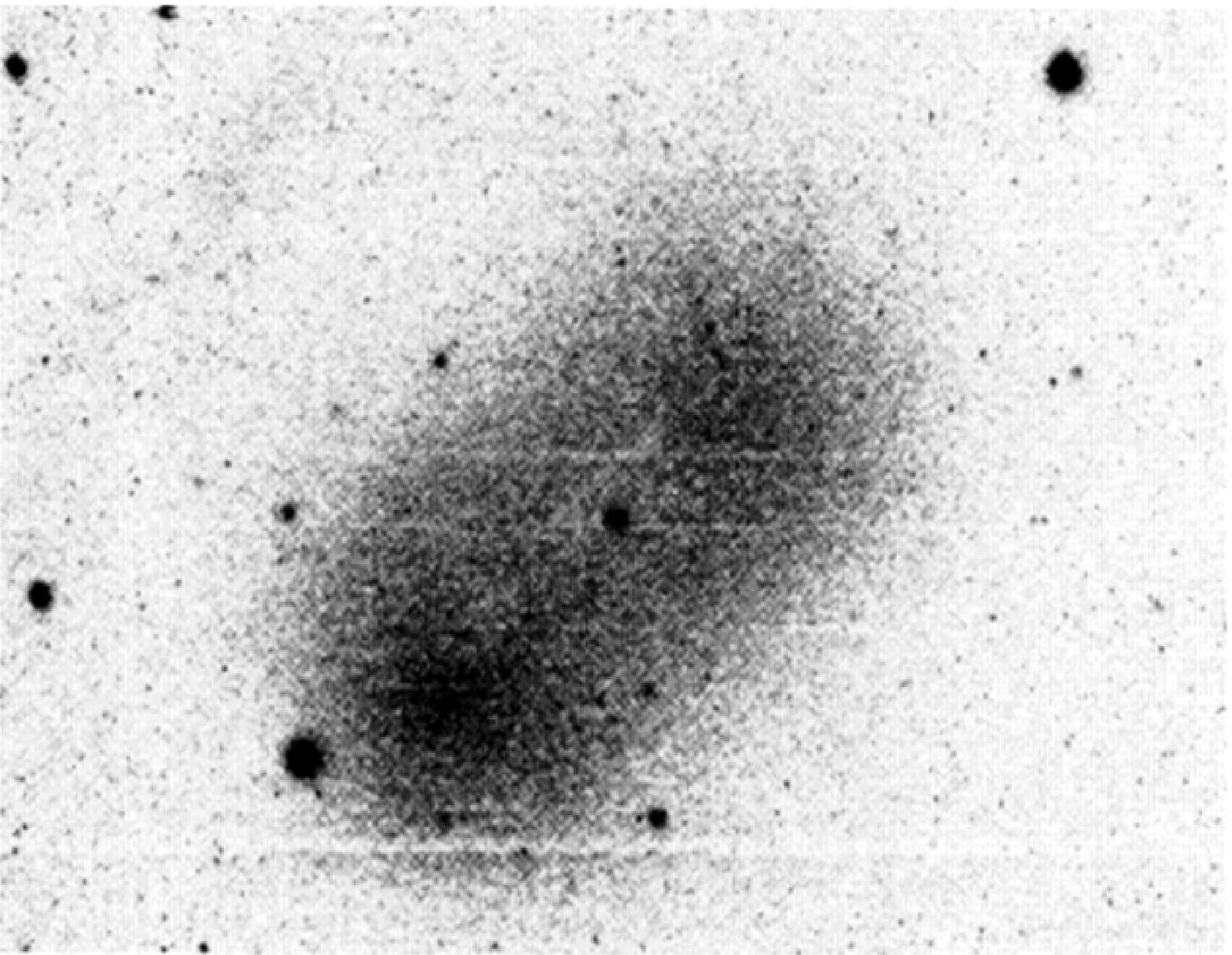}
\caption{{\bf Edge-on waists}: H~2-29 [top left], FOV$\sim$1', North is towards the top, East to the left (data from \citealt{Schwarz1992}; color image from the PN Image Catalogue: R,G,B=log(H$\alpha$), both, log[OIII]); A~63 [top right], FOV$\sim$1.5', North is towards the top, East to the left (from \citealt{Mitchell2007}, reproduced with permission; Ha+[NII]); HaTr~4 [bottom left], FOV$\sim$1',  North is towards the top, East to the left (from \citealt{Bond1990}, reproduced by permission of the AAS; [OIII]); A~65 [bottom right], FOV$\sim$4'x2.5', North is towards the top, East to the left (from \citealt{Bond1990}, reproduced by permission of the AAS; [OIII]).}
\label{fig:edge-on}
\end{figure*}

\begin{figure*}
\includegraphics{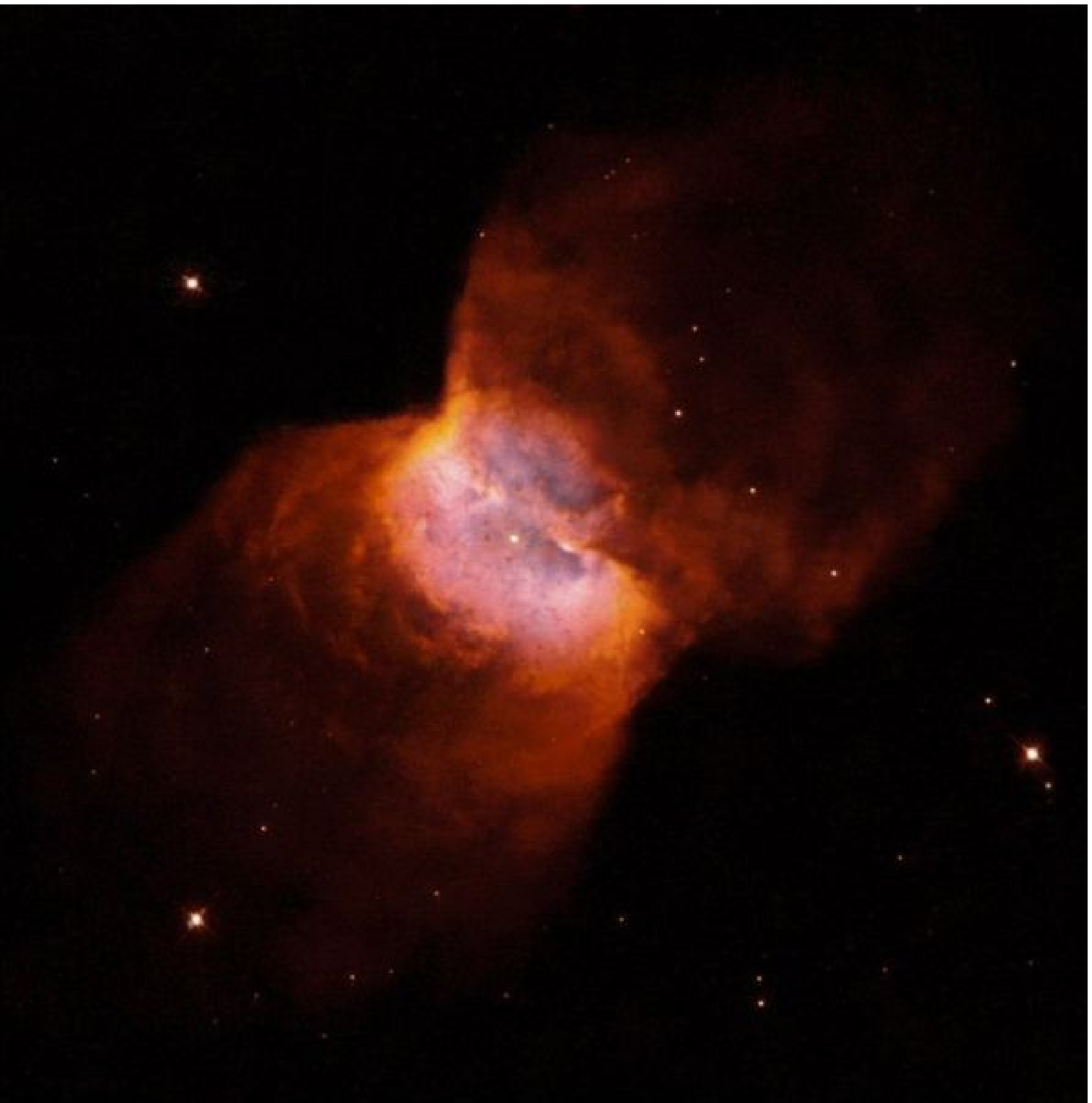}
\includegraphics{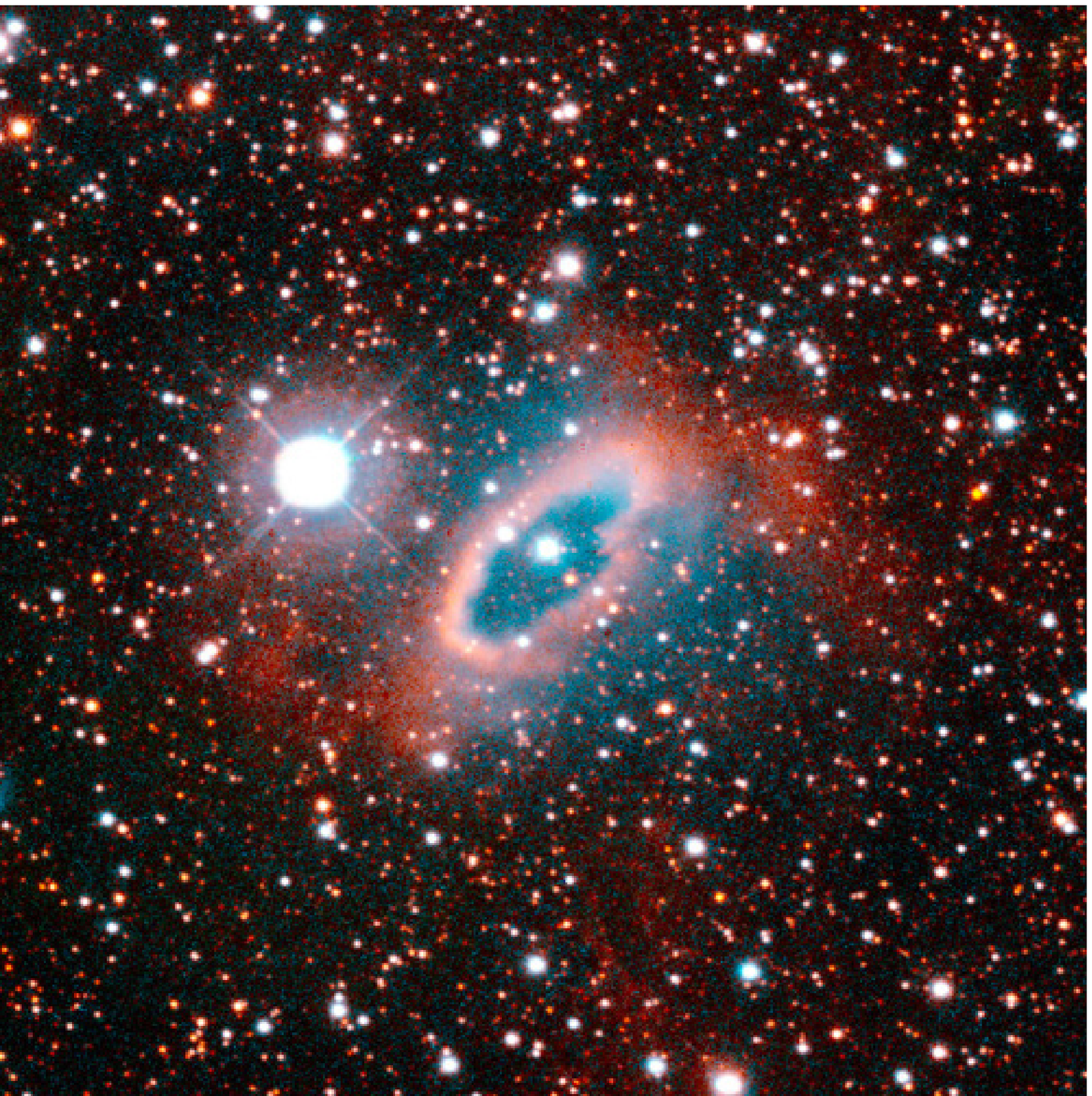}
\includegraphics{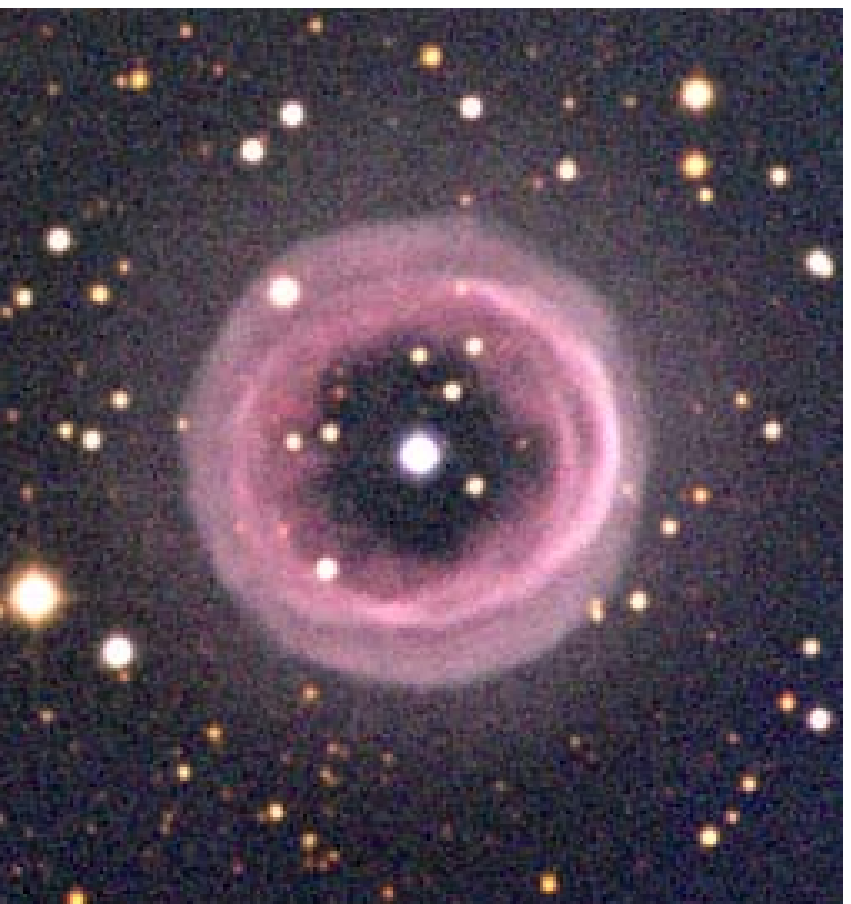}
\includegraphics{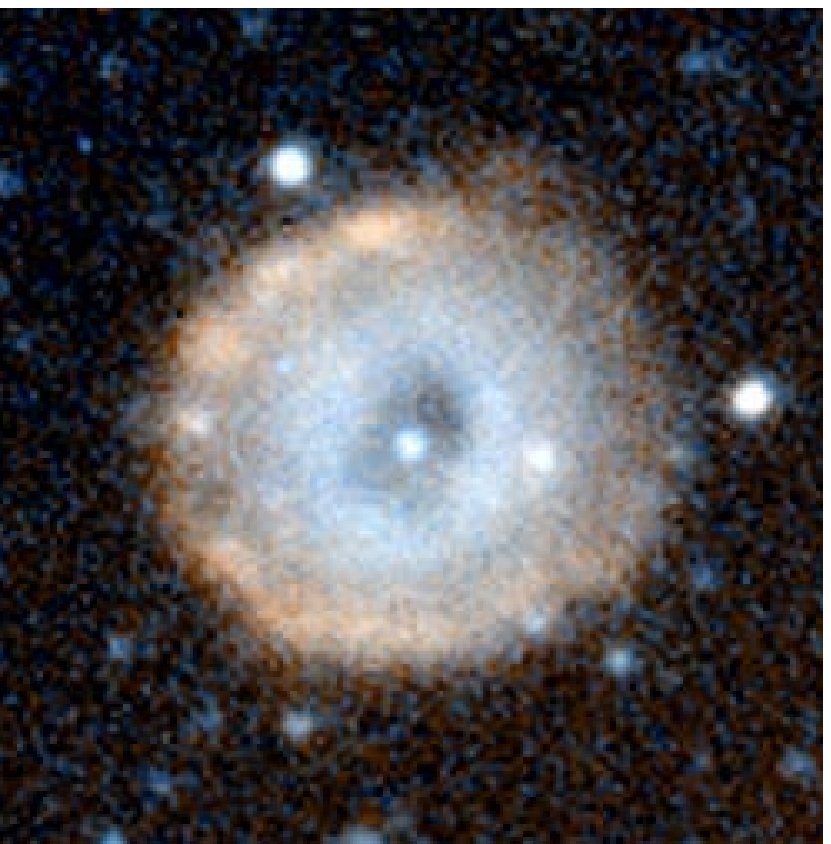}
\includegraphics{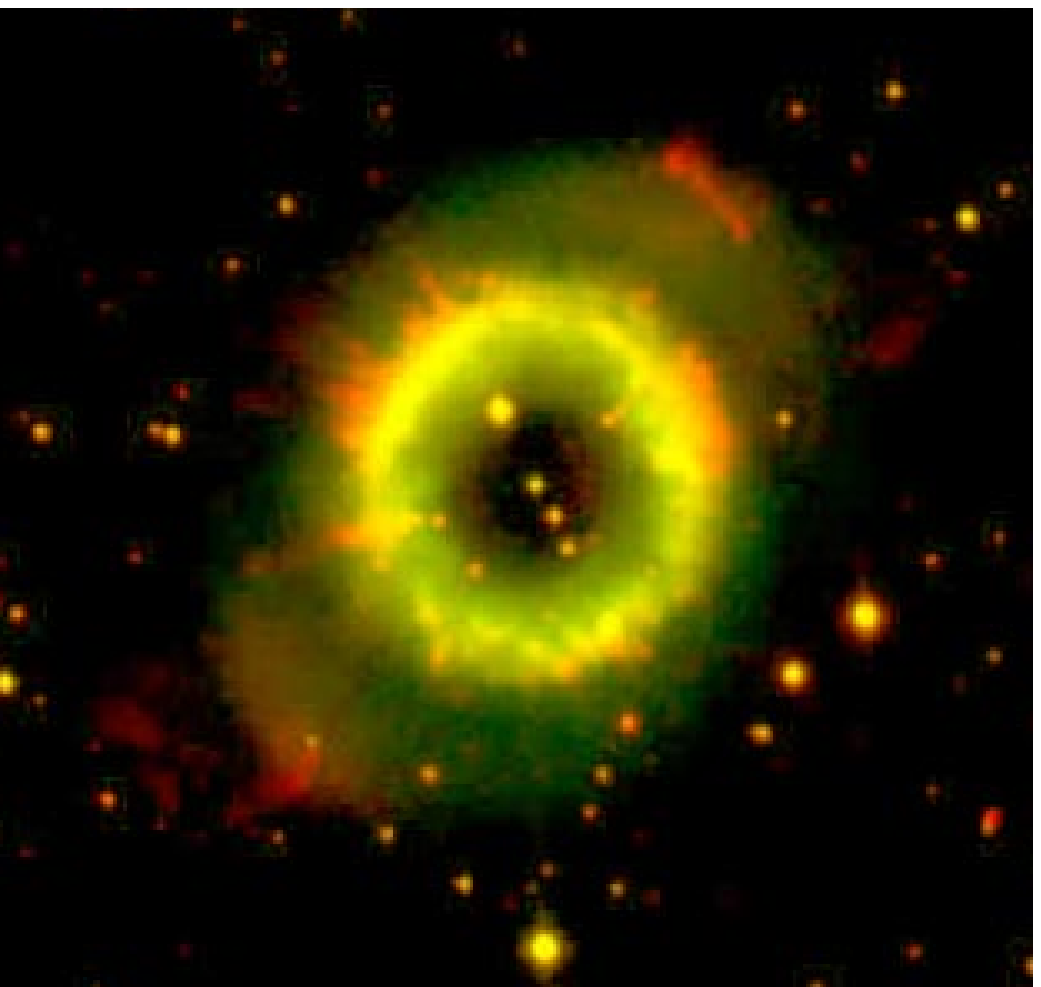}
\caption{{\bf Bipolar lobes}: NGC~2346 [top left], FOV=2.8', North is towards the top-right, East to the top-left (HST/WFPC2 image, color image assembled by The Hubble Heritage Team: R,G,B=[OIII], H$\alpha$,[NII]); SuWt~2 [top right], FOV$\sim$9', North is towards the top, East to the left  (CTIO/NOAO 1.5-m telescope image; color image from H. E. Bond and K. Exeter; R,G,B=H$\alpha$, both, [OIII]). This object has faint bipolar lobes protruding from an inclined ring. {\bf Rings}: Sp~1 [bottom left], FOV (width)$\sim$2', North is towards the top, East to the left (image from the Anglo-Australian Observatory, photograph by David Malin; R,G,B$\sim$B,V,H$\alpha$). This PN has a pole-on ring; Hf~2-2 [bottom center], FOV$\sim$1', North is towards the top, East to the left (data from \citealt{Schwarz1992}; color image from the PN Image Catalogue: R,G,B=log(H$\alpha$+[NII]), both, log[OIII]); this PN might have a ring, although the volume encircled by the ring is not evacuated; NGC~6337 [bottom right], FOV$\sim$2', North is towards the top, East to the left (data from \citealt{Corradi2000}; image courtesy of R. L. M. Corradi and the PN Image Catalogue; R,G,B=[NII],[OIII],null); this PN is an example of a thick ring with, possibly two lobes seen close to pole on, and a pair of knots.}
\label{fig:rings}
\end{figure*}

\begin{figure*}
\includegraphics{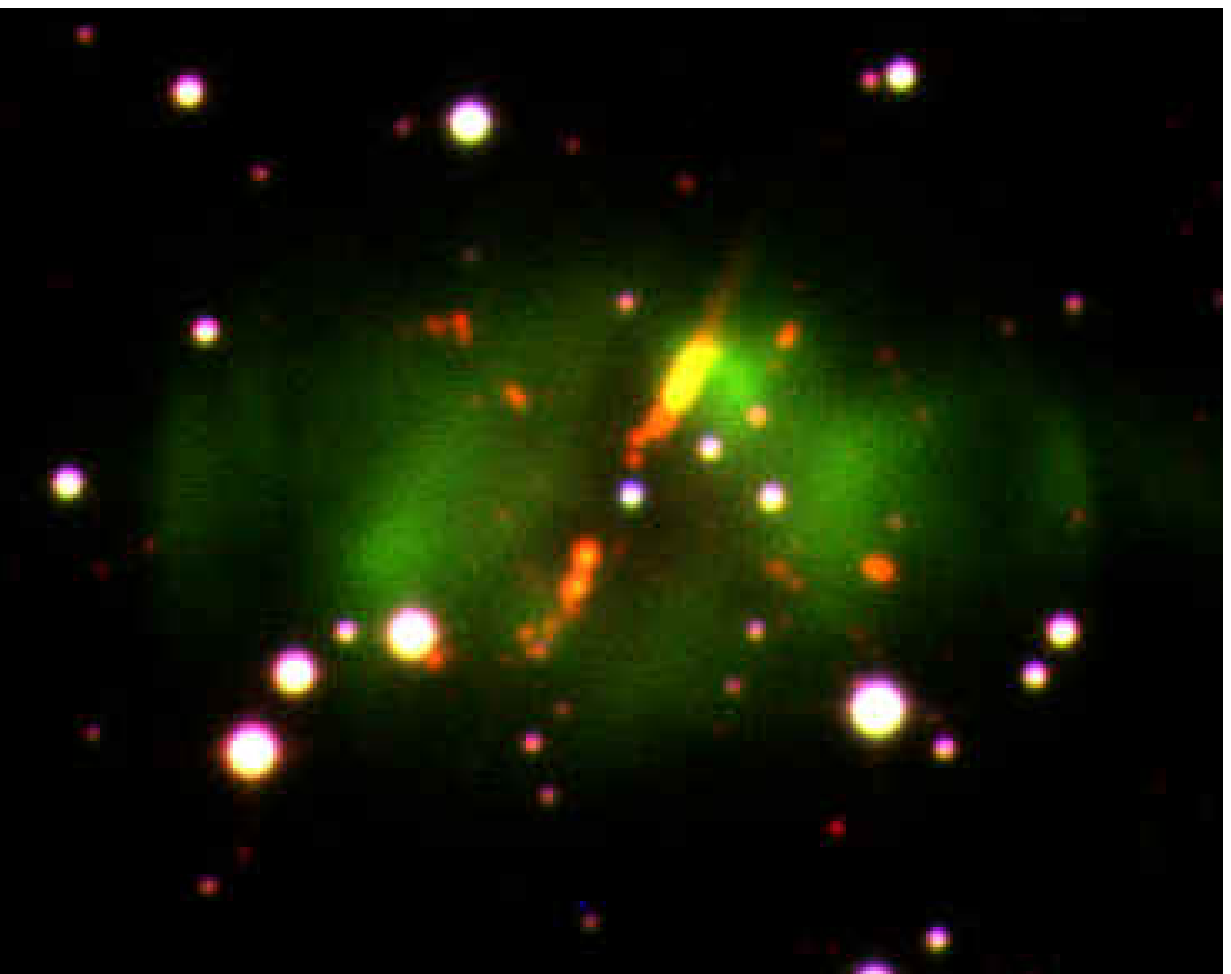}
\includegraphics{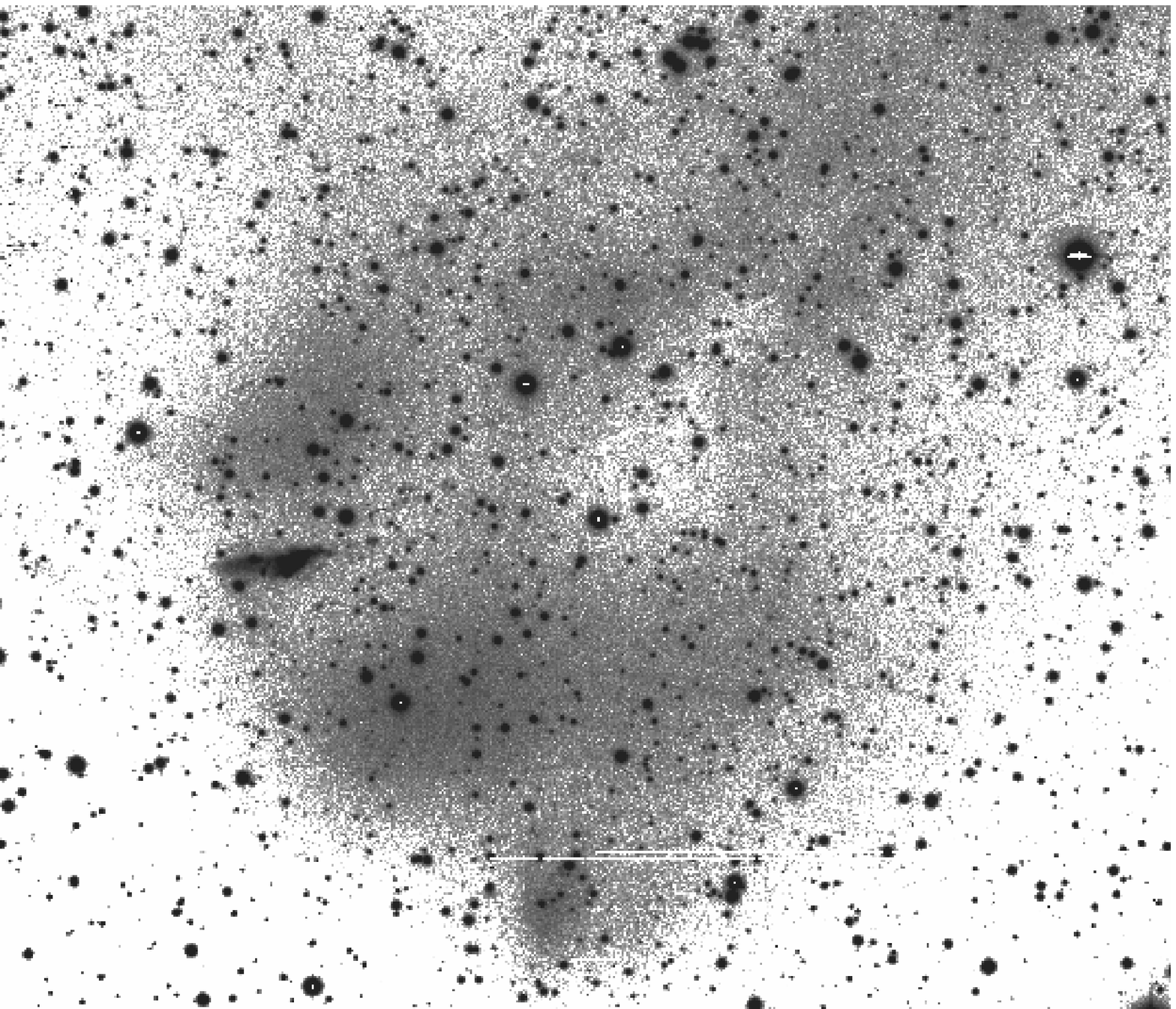}
\includegraphics{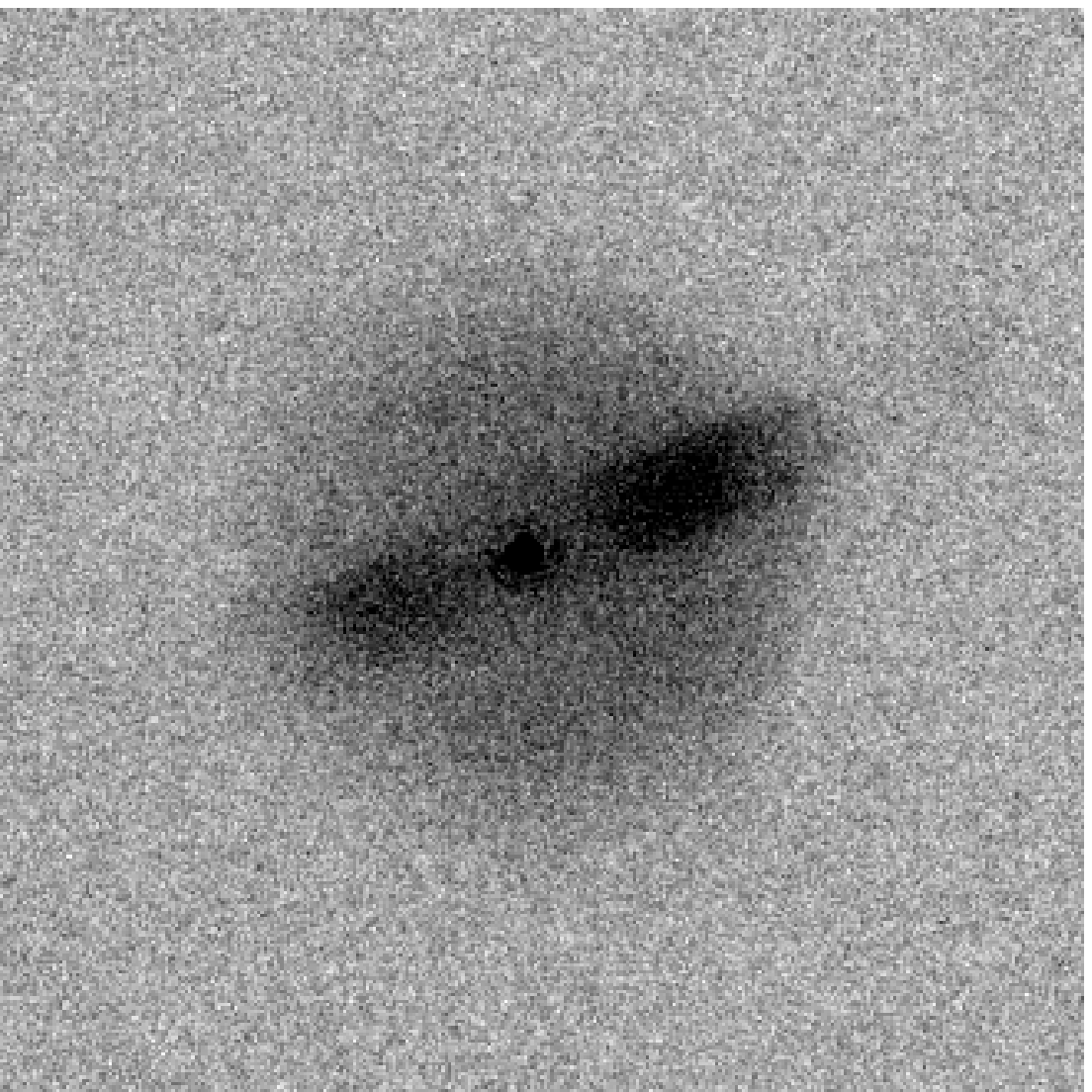}
\includegraphics{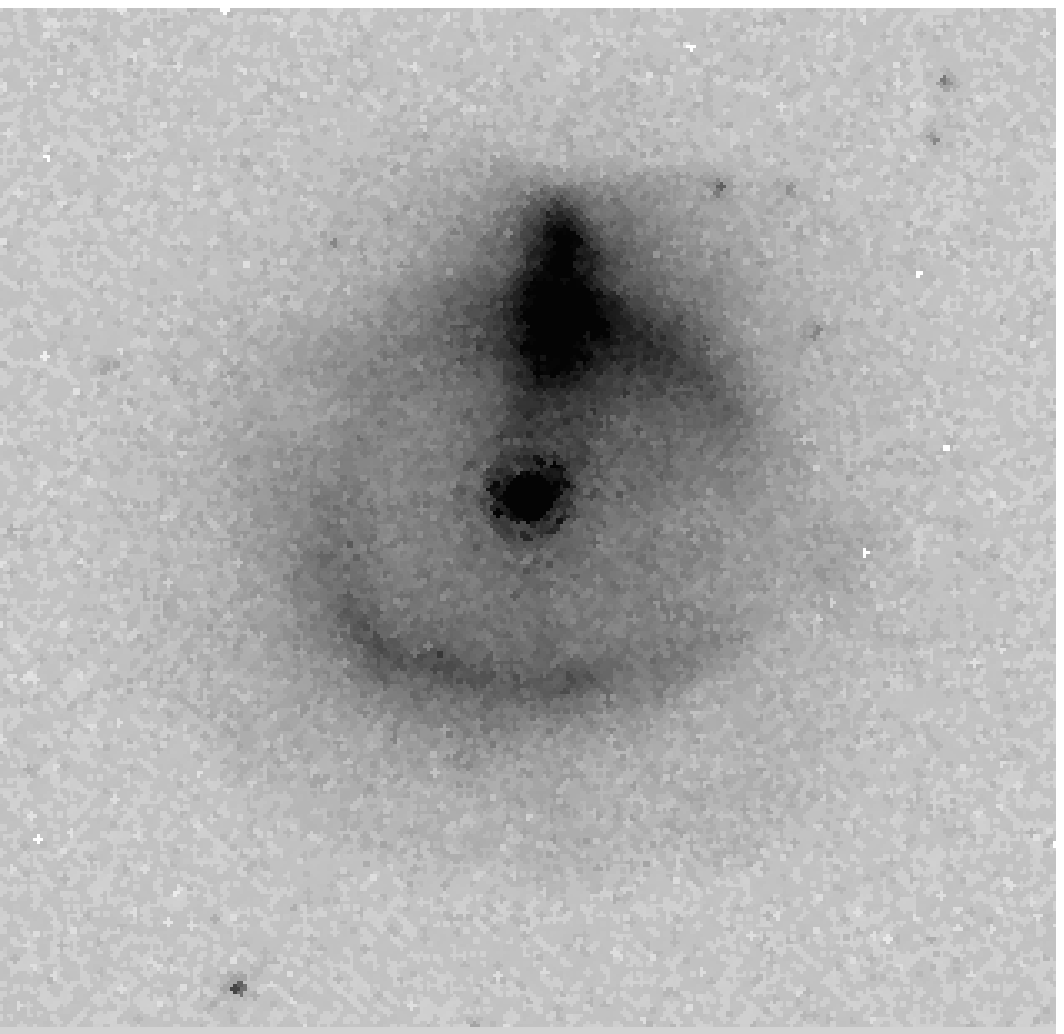}
\caption{{\bf Jets}: K~1-2 [leftmost], FOV$\sim$2.5'x2', North is towards the top, East to the left (data from \citealt{Corradi1999}; color image courtesy of R. Corradi, see also the PN Image Catalogue: R,G,B=[NII],log(H$\alpha$),--); HFG~1 [second from left], FOV$\sim$15'x10', North is towards the top, East to the left (unpublished image courtesy of R. Corradi, taken at the Isaac Newton Telescope in the light of H$\alpha$+[NII] ); PNG~135.9+55.9 [second from right], FOV$\sim$10'', North is towards the top, East to the left (HST/HRC H$\alpha$ image; see also \citealt{Napiwotzki2005}); M~2-29 [rightmost], FOV$\sim$6'', North is towards the top, East to the left (HST/WFPC2-PC H$\alpha$ image; see also \citet{Hadjuk2008}) -- this PN contains a strongly suspected triple system, but not yet confirmed.}
\label{fig:jets}
\end{figure*}

Here, we take another look at these objects and point out that many have structures that can be, at least qualitatively, explained as faded or projected bipolars. In Figs.~\ref{fig:edge-on} to \ref{fig:jets}, we show examples of edge-on  waists, rings, bipolar lobes and jets. We argue that these four feature types are expected from common envelope binary interactions that result in strong equatorial mass-loss \citep{Sandquist1998}, promoting the formation of bipolar PNe that, when faded, leave behind waists and rings. Jets, can form during or after the interaction if gas is accreted onto the companion or even onto the primary. 

Four PNe seem to be the thick waists of what was once bipolar in shape: A~63, H~2-29, HaTr~4 and BE~UMa (BE UMa: Bond, priv. comm). In addition, A~41 does not appear as a waist in the color image from the PI Image Catalogue, but it does look like a waist in the original black and white display of the same images \citep{Schwarz1992}, warning us that morphology is not only dictated by the filter and depth of the image, but also by the specific display choices. Another, A~65 also appears to be an almost-edge on torus, but once again the image reproduction is of low quality.  Five PNe have rings, but no lobes (Sp~1, M~2-19, M3-16 and, possibly, Hf~2-2 and NGC~6337). Two might have rings, A~65 and NGC~6026. A deep image of SuWt~2 shows faint lobes protruding from a ring which might indicate that all rings are thin waists of bipolar PNe \citep[see also][]{Zijlstra2007}. The exact relationship between thin rings and thick waists remains however unclear. It is possible that thin rings do not have enough mass to cause polar flows and are instead a result of axial mass-loss as opposed to its cause (Zijlstra, priv. comm.). 

Only 2 PNe are clearly bipolar, NGC~2346 and the newly discovered M~2-19, although, as we said, the ring-like SuWt~2 has faint bipolar lobes protruding from the ring.  Six PNe have one or two jets: A~63, M~3-16, HFG~1, K~1-2, PNG~135.9+55.9  and NGC~6337 (though for the last object the jets are more similar to a pair of bow shocks or knots). Of these, 3 PNe (HFG~1, K~1-2 and PNG~135.9+55.9, all shown in Fig.~\ref{fig:jets} along with M~2-29, the PN suspected of containing a triple system) have peculiar but similar morphologies, with a rounded or even ring structure onto which one or two transversal jets (or elongated structures) overlap. While the presence of jets is conceivable in the framework of a common envelope the overall shapes are difficult to interpret. The remaining 3 objects (A~46, DS~1 and Pe~1-9) are complex and/or faint PNe, difficult to classify (we do not classify them in Table~\ref{tab:knownbinaries}). Of the 21 PNe around binary central stars discovered by \citet{Miszalski2009} only 4 have published images clear enough for classification. 

The morphologies of the 25 PNe around composite spectrum and visual binary CSPNe, does not exhibit the remarkable features discussed above. The only noticeable characteristics of these PNe, is that most of them appear to have two approximately concentric structures, one brighter, on the inside, and one fainter, on the outside (e.g., NGC~2438, NGC~3132) and that there are two round PNe (A~30 [though this PN has non-round, hydrogen-deficient ejecta at its center] and A~33; but see points raised on hydrogen-deficiency and spherical PNe in \S~\ref{sec:additionalriddles}).

A full kinematic study is not easy in most of these objects because they tend to be faint (which is itself possibly an observational bias; \citealt{Bond1990}). For the eclipsing binary A~63, \citet{Mitchell2007} carried out a full kinematic study of the edge-on waist and the jets and solidly tied the PN to the eclipsing binary CSPN. A similar match is reached for K~1-2, where similar binary central star and PN inclinations have been determined by \citet{Exter2003b} and \citet{Corradi1999}, respectively. For NGC~6337, \citet{Corradi2000} concluded that the ring is an almost pole-on waist and that the two faint lobes are those of a bipolar nebula. This is in agreement with the study of the central binary carried out by \citet{Hillwig2006}. For Sp~1 the conclusion is also that the ring is the waist of a bipolar PN seen pole-on (Mitchell, private communications cited in \citealt{Zijlstra2007}). Rings and waists are therefore reasonably well associated to bipolarity. If so, as many as 14 out of 20 PNe around close binaries with adequate images may have a  bipolar morphology. PNe such as K~1-2, or PN~G135.6+55.9 do however remain enigmatic.

{\it Abundances.}  The PN C/O ratio and the abundance of s-process elements, gauge the amount of third dredge-up that took place in the precursor AGB star, since after every helium thermal pulse the convective layer extends downwards and dredges up carbon (the product of shell helium burning), as well as s-process elements. As a result, an AGB star is initially oxygen-rich (C/O$<$1 by number) but becomes progressively enriched in carbon, so that its envelope's C/O ratio eventually becomes larger than unity. The super-wind occurs at the end of the AGB phase and imparts to the PN its own C/O ratio and s-process element abundances. In addition, if the N/O number ratio is larger than 0.8, we call the PN Type I \citep{Kingsburgh1994}. This high N/O ratio is thought to point to a higher progenitor mass, since only the more massive progenitors ($M_{MS} \ga 4$~\msun) are hot enough at the bottom of their AGB convective envelope to burn carbon into nitrogen. As a result, PNe with a high N/O ratio tend to have a lower C/O ratio.  

If an AGB binary enters a common envelope interaction with its companion it quickly departs from the AGB with the entire envelope being ejected in the space of a few years \citep[e.g.,][]{Sandquist1998}. Statistically, the common envelope happens before the natural termination of the AGB so that PNe which are ejected common envelopes should, statistically, not be as carbon-enriched as those that terminated the AGB naturally \citep{Izzard2004,Izzard2006}. As we explained above, a low C/O ratio could also result from conversion of carbon to nitrogen and is therefore quite natural in type I PNe. One might therefore predict that PNe resulting from common envelopes, would have a systematically low C/O ratio but have any range of N/O ratios. Unfortunately very little data exist on the abundances of PNe around close binary CSPNe. 

The only confirmed close binaries with determined C/O {\it and} N/O values are Hf~2-2 (C/O=0.48 and N/O=0.38 from the optical recombination line analysis of \citet{Liu2006}) and NGC~2346 (C/O=0.35-0.49; \citealt{Rola1994,Kholtygin1998} and N/O=0.45; \citealt{Perinotto1991}). If to these we add NGC~2438 (which does not have a known period; C/O=0.57, N/O=0.42; \citealt{Kingsburgh1994}), and the suspected close binaries (\S~\ref{ssec:suspectedbinaries}) M~2-29 (C/O=0.69 and N/O=0.35; \citealt{Kholtygin1998,Webster1988}),  Hb~12 (C/O=0.3-0.52 and N/O=0.092; \citealt{Aller1994,Hyung1996,Perinotto1991}) and NGC~6302 (C/O=0.31-0.75 and N/O=1.65; \citealt{Zuckerman1986,Tsamis2004}), we see that all C/O values are below unity, and that only one of those (for NGC~6302) could be justified as deriving from carbon conversion to nitrogen in a type I PN. (We do not bring to bear here the abundances determined by \citet{Pollacco1997} for  the PNe with close binaries A~41, A~64, A~63 and A~65, because their values for the C/O ratios were derived via a mixed recombination and collisional line analysis, which is known to give inconsistent results; \citealt{Liu2006}). 

Regarding the s-process elements, \citet{Sterling2008} found that the mean abundance of  selenium and krypton in stars suspected of being binaries (though none of the stars in their list has been confirmed) is indeed smaller than for the average of other stars.

Finally we should point out a set of correlations between morphology, PN kinematics and abundances which are difficult to explain under any scenario. In the Galaxy, bipolar PNe have a smaller scale height than non bipolar ones, larger expansion velocities and higher N/O ratios \citep{Corradi1995}. In addition, using a Large Magellanic Cloud (LMC) sample, \citet{Stanghellini2007} determined that the only PNe to have a low C/O ratio are bipolar, which would be consistent with their Type~I nature.  All these characteristics point to bipolarity being due to a higher progenitor mass\footnote{We should point out that \citet{Villaver2007} threw a spanner in the works by finding that, in the LMC, bipolar and elliptical PNe share the same mean CSPN mass, contrary to expectations that bipolars should have a higher mass. However, mass-determinations are much more uncertain than either abundances or scale-height measurements.}. In the Single Star Paradigm this would point to more massive (single) star being able to produce bipolar PNe by some mechanism.
For the Binary Hypothesis, \citet{Soker1998} proposes that primaries that undergo a common envelope on the AGB (resulting in a bipolar nebula)  would most frequently have higher mass. This would result from the fact that lower mass primaries have relatively larger RGB radii and interact more readily on the RGB. These lower mass stars would therefore never ascend the AGB nor would they make PNe. However, taking a look at Table~\ref{tab:knownbinaries} we see that the masses determined for  primaries in close binaries are not particularly high, nor are the N/O ratios of their PNe (for the few objects with data).

\subsection{Suspected binary central stars}
\label{ssec:suspectedbinaries}

Some CSPNe can be rightly suspected of being binaries. These include, primarily, PNe with cool CSPNe which are unable to ionize the PN and a few other PNe for which current observations indicate the presence of a companion, but are insufficient or of insufficient quality to confirm the system as a binary. Finally, although we do not want to use morphology to prove binarity, there is a subset of PN morphologies which are extremely likely to harbor or have harbored a binary system. These are the PNe with a strong point symmetry, those where the CSPN is visibly offset from the center \citep{Soker1998b} and finally, those exhibiting jets or jet-like features \citep{Soker1994b}.

{\it Cool CSPNe.} In Table~\ref{tab:coolcentralstars} we list CSPNe with effective temperatures (measured or deduced from their spectral types) too low to ionize their own PN. This is usually taken as an indication that an undetected hot companion is present.   When these cool CSPNe are evolved stars, it is possible that the system is in a symbiotic nebula instead of a PN (see Sec.~\ref{ssec:symbiotics}).

Another explanation for cool spectral types is that the star is in a post-AGB phase and is just about to ionize its PN. The PN might in fact be a low-ionization object, where the emission lines are  formed because of collisional ionization due to the passage of shocks, rather than photo-ionization. Any time a PN is detected (by the presence of emission lines) around a B-type supergiant this is a distinct possibility. An example of this kind of object could be IRAS~19336-0400 \citep{Pereira2007}, although even for this object the presence of a hot companion has been suggested \citep{VandeSteene1996}. If a hot companion is present, the object could of course be a symbiotic binary.

Finally when a CSPN has a cool spectral type, another possibility is a chance superposition of a PN with a cool star. This is likely to be  the case for M~1-44, for which  the K2 III central star \citep{Lutz1977}, was later determined to be a chance alignment by \citet{Pereira2004} from radial velocity measurements.

{\it PNe with jets, point symmetry or with off-center CSPNe.} PNe with point symmetry (e.g., IC~4634 \citep{Hyung1999}, PC~19 \citep{Guerrero1999}, or He~2-186 \citep{Corradi2000}) constitute a few percent \citep[][and private communication]{Balick2008} of all PNe and are extremely hard to explain with single stars which, at most, would impart an axi-symmetry. However, jets from a binary companion which accretes material from the primary can precess or ``wobble" and impart the observed point symmetry \citep{Livio1996,Soker1997}. It is therefore likely that PNe with point symmetry did indeed derive from a binary interaction. 

When jets or jet-like structures are observed in PNe, a binary interaction is usually suspected, since it is easier to produce and collimate jets in binary configuration. Jets can be produced by a disk around the companion or one around the CSPN itself. Eleven percent of a sample of 618 PNe analyzed by \citet[][and private communication; e.g., IC~4593]{Balick2008} shows the presence of ansae or jets. Yet other PNe have less collimated structures which might or might not be jets and over whose nature people are still arguing.  

\begin{figure}
\includegraphics{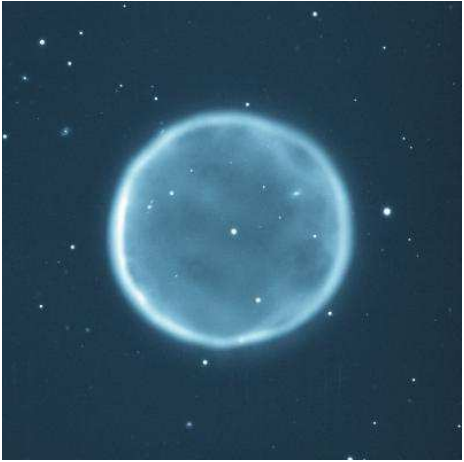}
\caption{The spherical PN A~39. FOV$\sim$335''x328'', North is towards the top, East to the left (WIYN; image data from  \citet{Jacoby2001}, image courtesy of  G. H. Jacoby; [OIII]).}
\label{fig:abell39}
\end{figure}

Finally, when motion through the ISM cannot explain an offset of the CSPN from the geometric center of the PN, a wide companion might \citep{Soker1997}. Interestingly, a good example of this is A~39 (Fig.~\ref{fig:abell39}), an otherwise spherical PN where the CSPN is displaced by 2\arcsec\ from the center of the PN in the opposite direction from what would be expected for its motion through the ISM, as deduced by the PN surface brightness \citep{Jacoby2001}.

{\it Other suspected binaries.} Certain PN and CSPN characteristics point to an undetected companion. The CSPN of Ton~320 has IR excess and RV variability, but more data is needed to confirm it \citep{Zuckerman1991,Good2005}. NGC~6853 might have a companion with M spectral type at 1\arcsec, but a better estimate of its distance is needed to insure that the cool star is associated with the PN \citep{Zuckerman1991}. The bipolar PN NGC~6302 is suspected of being a close binary based on a kinematic study of its molecular torus \citep{Peretto2007}. From radio observations, \citet{Bains2004} strongly suspect the bipolar PN Mz~3 to have a binary companion (but this object might also be a symbiotic binary; \citealt{Schmeja2003}). 

Large dusty disks detected through various techniques (e.g., because of light occultation and reflection, mid-IR excess or semi-periodic light variability) are also a likely indication of binarity.
The confirmed binary CSPN of NGC~2438 has a disk detected because of an IR excess. Another confirmed binary, that inside PN NGC~2346 has a disk detected because of light variability. Although, to connect disks to binarity a full theory of binary-induced dusty torus formation is needed \citep[see, e.g.,][]{Peretto2007,Nordhaus2007}, one can keep CSPNe with dusty disks in the list of suspects. Examples of CSPNe suspected of being (or having been) binaries could therefore be the [WC10] CSPN CPD-56~8032 that has a dusty disk detected because of central star light occultation and semi-periodict variability \citep{DeMarco2002b}, or M~2-29, that has an almost identical behavior to NGC~2346,  and that is suspected of containing a triple system with a 23 day and an 18 year periods \citep{Hadjuk2008}. A last example is Hb~12 that has semi-periodic eclipses that might to be caused by orbiting dust \citep{Hsia2006}. 

\citet{Rodriguez2001} detected very high density, unresolved circumstellar material around the CSPNe of He~2-428 and M~1-91, which they interpreted as deriving from some kind of accretion process (this is also the case for EGB~6, but in this object the dense cloud of material has been resolved and it is around the visual companion to the CSPN; \citealt{Bond1993}). \citet{GonzalezPerez2003} detected very fast variability of the hot star in NGC~246 (that is hydrogen-deficient and has a wide binary companion -- Table~\ref{tab:widebinaries}) which they attribute to a binary interaction with a putative close companion.

Additionally, specific chemical anomalies in the PN gas might also point to present or past binarity. For instance \citet{Otsuka2008} has argued that the high fluorine abundance in BoBn~1 groups this object with the carbon-enhanced, metal-poor stars \citep[CEMP;][]{Aoki2007}, most of which are thought to derive from the interaction with a more massive companion that was on the AGB in the past. Finally, if a CSPN has characteristics explained as a past mass transfer episode, one can also postulate that it is or was a binary. This is the case for the barium CSPN of WeBo~1 \citep{Bond2003}.


\section{Predictions and tests of\\ the Binary Hypothesis}
\label{sec:predictionsandtests}

The Binary Hypothesis simply states that a majority of PNe has formed through a binary interaction channel. Here we quantify this statement by reviewing population synthesis and other work that predict the CSPN binary fractions and period distributions. We also carry forward the consequences of these predictions and determine which other testable predictions derive from them. 

Brown dwarfs and planets are excluded from these predictions, mainly because population synthesis work has not yet accounted for their presence. In addition, even if we were to attempt an accounting of such low mass companions, we still do not know what their actual frequency is around main sequence stars (see \S~\ref{ssec:mainsequence}), not to mention what their longevity would be around an evolving expanding giant.

\subsection{The binary fraction and period \\distribution (with stellar-mass \\companions)}
\label{ssec:fracanddist}
 
Population synthesis studies take a zero age main sequence population, a fraction of which consists of binary stars with a given period distribution, and evolve it by following the rules of stellar evolution and binary interactions. Starting with a main sequence population in which 50\% of the stars are in binaries with periods shorter than 100 years, \citet{Han1995} predicted that (38$\pm$4)\% of all PN have been affected by binary interactions.   The post-common envelope evolutionary channel accounts for a third of this fraction ($\sim$13\%), with the other two thirds accounted for by mergers and very weak, wider binary interactions, in equal proportions. 

\citet{Moe2006} used a population synthesis technique to predict the total number of PNe with radii smaller than $\sim$0.9~pc, that reside in the Galaxy today: 46\,000$\pm$22\,000. They compared this predication to  the observationally-based estimate of the PN Galactic population with radii smaller than $\sim$0.9~pc: 8000$\pm$2000 (\citealt{Jacoby1980}\footnote{This number has a complex origin explained in full in \citet{Moe2006}, but to justify it here, suffice it to say that if the Galaxy had a much larger PN population than this, it would be very peculiar compared to other galaxies in the Local Group. Larger estimates are seen in the literature, however recent work on the local PN sample by \citet{Frew2008b} has the total Galactic PN population with radius larger than 0.9~pc at $\sim$11\,000 objects, in agreement with the assessment of \citet{Jacoby1980}. A full discussion of the different estimates can be found in Moe \& De Marco (in prep.).}). The prediction is larger than the observation by a factor of 6, at the 3~$\sigma$ level. They concluded that this discrepancy could be alleviated if only a fraction of stars currently thought to produce PNe, actually do. Such a subset could be the interacting binaries.  

Using the predictions of \citet{Han1995} and \citet{Moe2006} we can estimate the fraction of stars that went through a common envelope interaction in the {\it Binary Hypothesis}. We take the $\sim$13\% fraction of all systems that went through a common envelope \citep{Han1995} and use it in combination with the total Galactic PN population size predicted by \citet[][46\,000$\pm$22\,000]{Moe2006}  to determine that about 6000 PNe in the Galaxy today ($46\,000 \times 0.13$) have CSPNe that went through a common envelope interaction and are still in binaries. This means that if only common envelope interactions were able to produce PNe, there should be 6000 PNe in the Galaxy today. Comparing this number to the 8000$\pm$2000 PNe in the Galaxy today, argues that $\sim$75\% of the Galactic PNe can be ascribed to a common envelope interaction. 

In conclusion, the post-common envelope CSPN fraction predicted by the Binary Hypothesis ($\sim$75\%) is  much higher than the observed post-common envelope CSPN binary fraction (12-21\%; \S~\ref{ssec:binarycentralstarsearches}). On the other hand the post-common envelope binary fraction predicted by the Single Star Paradigm (13\%) is in line with the observation. 
In the Binary Hypothesis the other 25\% of CSPNe (2000 objects) have originated via other mechanisms, such as weaker binary interactions, or single stars mechanisms. The remaining 38\,000 (46\,000--8000) objects did not make a PN at all and are now ``naked" central stars (see \S~\ref{ssec:additionalpredictions}).  

Taken at face value, the argument above would suggest that the Binary Hypothesis predicts too many (75\%) post-common envelope binaries and is discrepant with the observations (12-21\%). This would be so, if we firmly believed that the predicted and observed fractions strictly describe {\it only} post-common envelope CSPNe. However, the prediction is likely to include also systems that went through a strong interaction but avoided the common envelope, resulting in binaries with periods longer than can be detected with the irradiation technique. We know that such binaries must exist since  30\% of the immediate progenitors of CSPNe, the post-AGB stars, are in binaries with periods 100-1500 days (\S~\ref{ssec:pagb}). These types of interactions are not accounted for by population synthesis calculations which likely count them among the common envelopes, inflating their post-common envelope prediction. It follows that the population prediction includes a type of binaries which could not have been detected by the observations. We wonder whether the intermediate period binaries are enough to bring the observed and predicted post-common envelope binary fractions in the case of the Binary Hypothesis in better agreement. 


\citet{Frew2008} analyzed 32 objects with 2MASS or DENIS near-IR photometry and deduced that $>$53\% of PNe have a cool companion. This number is a lower limit because detections are limited to spectral types brighter than M0V-M8V, depending on the brightness of the hot CSPN. However, this number is also an upper limit because it is likely to include some binary CSPNe that are not resolved by 2MASS or DENIS, but that are still too wide to have interacted. Improved near-IR surveys are a promising avenue to determine the binary fraction for systems with separations out to a few tens to $\sim$100~AU, in particular in combination with high resolution imaging that can exclude the wider systems from the accounting. 


\subsection{Additional predictions}
\label{ssec:additionalpredictions}

{\it PN birthrate and population density}. \citet{Moe2006} predict a PN birthrate density of 1.1$\times$10$^{-12}$~PN~pc$^{-3}$~yr$^{-1}$ if single stars and binaries (Single Star Paradigm) produce PNe (which is close to the prediction of \citet{Liebert2005} for WDs). In the Binary Hypothesis this birthrate density should be about $\sim$6 times lower, or $\sim$0.2$\times$10$^{-12}$~PN~pc$^{-3}$~yr$^{-1}$ \citep[][see also \S~\ref{ssec:fracanddist}]{Moe2006}. This is substantially lower than current determinations \citep[e.g.,][]{Phillips2002,Frew2008b}, resting on the distance scale.  By predicting a smaller PN density, the Binary Hypothesis therefore predicts a lengthening of the distance scale. 

{\it PNe in GCs}. Single star theory predicts that there should be about a dozen hot post-AGB stars in GCs \citep[e.g.,][]{Moehler2001,Zinn1974}, which is about the observed number. However, none of these should have a PN because their low mass translates into long evolutionary times;  by the time they are hot enough to ionize the PN, the PN gas has dispersed. \citet{Jacoby1997} showed that there are 4 PNe in the Galactic GC system, discrepant with theory at the $\sim$3$\sigma$ level. The Binary Hypothesis predicts instead that there should be a few PNe in GCs because the common envelope can shorten the transition time between the AGB and the phase when the star has a temperature sufficient to ionize the PN (Moe \& De~Marco , in prep.; see also \S~\ref{sec:additionalriddles} for an additional justification of the presence of PNe in GCs.)

{\it Naked central stars}. If only binaries can make PNe, it would follow that there would be a population of hot post-AGB stars that do not have a PN in virtue of the fact that they are either single or are in binaries too wide to have interacted. According to the predictions of \citet{Moe2006} there should be $\sim$6 times as many naked central stars as central stars with a PN.   \citet{Soker2005}  put forward the hypothesis that a single star might only be able to make a spherical and under-luminous PN. It is possible that these naked central stars are currently known as sub-dwarf O stars. To distinguish them from hot post-RGB objects that also have an sdO spectrum, one would have to carry out a stellar analysis to determine $\log g$ and $T_{eff}$ so as to compare these values with stellar evolutionary tracks (e.g., \citealt{Mendez1988b}). These post-AGB sdO stars would be far inferior in number compared to post-RGB sdO stars, due to the faster post-AGB evolutionary time-scales, so it might be very hard to identify them.

{\it PN abundances.} As we have pointed out in Sec.~\ref{ssec:morphologykinematics}, the C/O ratio of PNe around post-common envelope CSPNe should be statistically smaller than for PNe that are not ejected by a common envelope interaction. However, this ratio is likely to depend also on the mass of the progenitor. So, till we have understood the relationship between PN C/O and N/O ratios, morphology and scale height, it is difficult to make a clear prediction as to the expected abundances of PNe around binary CSPNe.  

{\it Wide binary CSPNe.}  In the Single Star Paradigm the period (separation) distribution of the wide binaries should reflect that of the main sequence (making allowance for a few changes due to angular momentum conservation and mass-loss). In the Binary Hypothesis the majority of these wide binaries should be triples where the secondary is close to the primary and the tertiary is a wide companion (see Sec.~\ref{sssec:widebinaries} for a longer discussion).

{\it X-rays from CSPNe.} We expect that the coronae of spun up companions in post-common envelope binaries would emit hard, line-dominated X-rays, while the disks around any of the two stars in a close binary might emit hard, continuum-dominated X-rays \citep{Soker2002,Jeffries1996}. Observations have already partly borne out these predictions, although, the number of known binaries with X-ray observations is small (the binary CSPN of LoTr~5 coincides with an X-ray point source consistent with a coronally active companion, while those of the PNe A~63 and LoTr~1 have not been detected in X-rays, but that could be due to their very high reddenings; \citealt{Apparao1992}). There are, however, some PNe whose CSPNe, although not known to be binaries, are detected in hard X-rays, which could be an argument for a close companion \citep[NGC~7293, NGC~6543;][]{Guerrero2001}. More X-ray observations of known binary CSPNe are needed to determine what to expect. 

{\it Magnetic fields incidence.} Observations of magnetic fields in AGB and post-AGB stars as well as in pre-PN and PN central stars can aid in determining which circumstances gave rise to them. For instance \citet{Jordan2005} detected magnetic fields in four CSPNe. \citet{Vlemmings2006} and \citet{Etoka2004} have detected magnetic fields in evolved AGB stars (thought to be  in transition between the AGB and post-AGB phases). In the case of W43A, the magnetic field seems to be collimating a high velocity jet emanating from the center of the circumstellar envelope. \citet{Bains2003,Bains2004} have detected milli-Gauss magnetic fields in pre-PN with large scale structures.  As explained in Sec.~\ref{ssec:secondaryshapingagents} magnetic fields are likely to be sustained because of the action of a companion, that is either still present or has merged. As theoretical work continues to predict the relationships between fields and binary configurations, observations of magnetic fields in binary systems with well determined parameters will provide an observational check of these predictions.


\section{Central stars in the context of their progenitors and progeny}
\label{sec:progenitorsandprogeny}

In this section we will present CSPNe within the context of their progenitors (AGB and post-AGB stars with and without a pre-PN) and progeny (WDs), starting with a short summary of the multiplicity frequency, including substellar companions, for main sequence stars that evolve to the PN stage. 


\subsection{Main sequence star multiplicity}
\label{ssec:mainsequence}

To determine  to what extent the current PN population is a result of binary interactions, we need to know how frequently main sequence stars with spectral type G, F and A, are in binaries with separations between a few and $\sim$100~AU (smaller separations than a few AU will result in a strong interaction during the RGB phase, which will most likely preclude the star from ever reaching the AGB and PN phases). In particular this information  is needed for main sequence stars with M$\sim$1.2~\msun\ with solar metallicty, because they form the bulk of today's CSPN population \citep{Moe2006}. 

For stellar companions, the binary fraction is reasonably well known down to $M_2 = 0.1-0.5$~\msun\ \citep{Duquennoy1991} and this information has been used in the population synthesis efforts described in \S~\ref{sec:predictionsandtests}. The main sequence binary fraction with brown dwarf companions is likely to be $<$1\% for systems with a few AU separation \citep[the Brown Dwarf Desert;][]{Grether2006} but rise to a few percent for brown dwarfs farther out \citep{Metchev2008}. The fraction of solar-type stars with jupiter-class planets (0.3-10 Jupiter masses) at $<$20~AU from the mother star is about 17-20\% \citep{Cumming2008}, although these numbers are still derived by extrapolating survey biases and should be regarded as provisional. We also know that the fraction of stars with planets is metallicity dependent \citep{Fischer2005}, with a steep gradient between 3\% for $-0.5>$[Fe/H]$>$0.0 and 25\% for [Fe/H]$>$0.3. The planetary frequency is also likely to be correlated with the primary mass \citep{Johnson2007}. As time goes by, RV surveys will extend their maximum detected separations and the fraction of stars with planets could increase (although it does already appear that the number of planets beyond 75~AU is low; \citealt{McCarthy2004}). 

To assess the impact of brown dwarfs and planets on the PN population we need to (i) have a better estimate of the fraction of stars with planets and brown dwarfs as a function of metallicity, primary mass and separation so as to integrate these results in population synthesis work and (ii) to understand what effect such low mass companions can have when they interact with AGB stars.

\subsection{AGB stars}
\label{ssec:agbs}

What we want to know about AGB stars is how the super-wind is triggered and why they change mass-loss geometry. If AGB star progenitors are more massive than $\sim$2.5~\msun, it is possible that a simple Reimers-type mass-loss \citep{Reimers1975} can eject, at the end of the AGB phase, the required amount of mass in a relatively short time. However, most PNe, even in the younger Galactic thin disk, derive on average from a progenitor less massive than that \citep{Moe2006}. As a result we are still in need of a mechanism that can initiate the super-wind for the bulk of the population. It is possible that the mass necessary for a single star to initiate its own super-wind might be chemistry and metallicity dependant. If so, carbon-rich AGB stars would have a greater ease in promoting the super-wind (and could do so at lower mass) than oxygen-rich ones, and Magellanic Clouds stars would always find it more difficult to promote the super-wind and would only do so at larger mass \citep{Lagadec2008}. On the other hand, common envelopes and other types of close binary interactions have the capability of initiating a super-wind phase and remove the AGB envelope.

For a long time we thought that the AGB mass-loss geometry was spherical \citep[][and references therein]{Olofsson1999} with very few systems deviating from sphericity \citep[e.g., V~Hya, X~Her;][]{Kahane1996,Kahane1996b}. The change to axi- or multi-symmetric mass-loss happened at some point during the {\it post-AGB} evolution. Now we are finding that {\it many} AGB stars already possess a non-spherical mass-loss \citep{Castro-Carrizo2007}. Future work will reveal how many and what types of AGB stars have non spherical mass-loss and the relationship between asymmetries in mass-loss and the presence of a binary companion.

{\it AGB binaries.} AGB star binaries of interest to the current problem are those where a companion is close enough so that Roche Lobe overflow, tidal capture or wind accretion are likely to take place upon further expansion of the stellar envelope. However, the difficulty of detecting directly faint companions in the proximity of very bright, windy and dusty AGB stars has limited the number of systems known (see also \citet{Jorissen2008} for a review of how to detect binary AGB stars). RV surveys \citep{deMedeiros1999} have shown that of 1500 F-K IV-II stars, 11-24\% are spectroscopic binaries, while for the KII stars alone, which are more likely to be on the AGB, the fraction is 7-18\%. 

Some AGB stars in binaries have been known for a while (most notably, Mira; \citealt{Karovska1997,Wood2006}), others are suspected from secondary indicators (e.g., BM Gem; \citealt{Izumiura2008}). Recently \citet{Sahai2008} has demonstrated that a large majority of AGB stars with Hipparcos astrometry containing a ``multiplicity" flag, do actually contain a hot companion. It therefore appears that we are closing in on a full characterization of AGB binarity.
  
In the LMC, where it is easier to study complete populations, we know that photometrically-variable giants with Long Secondary Periods \citep[sequence D in, e.g.,][]{Derekas2006} are mostly AGB stars (as opposed to RGB stars). Sequence D comprises 25\% of all AGB stars studied in this way.  There are strong arguments that ascribe the long secondary periods to semi-detached binaries with stellar and sub-stellar companions \citep[e.g.,][]{Soszynski2004}; others argue that such an interpretation would lead to an unrealistically high AGB binary fraction, but the lack of a better explanation for the data has left the binary model standing. In support of the binary interpretation we note (see also Sec.~\ref{ssec:pagb}) that 30\% of all {\it post}-AGB stars are in {\it spectroscopic} binaries and this too is a very large (and similar) number, though it is hard to say whether the LMC AGB population polled by that survey is compatible with the Galactic post-AGB population polled by \citet{vanWinckel2003}.

\subsection{Post-AGB stars (with and without\\ pre-PNe)}
\label{ssec:pagb}

Post-AGB stars are the immediate progeny of AGB stars, after the super-wind has depleted the AGB envelope and the star   restructures by shrinking and heating. The post-AGB phase is technically over when the star is hot enough to ionize the PN ($\sim$25\,000~K), by which time we call the star a CSPN (or pre-WD). Post-AGB stars are found either with no circumstellar nebula or in the middle of a pre-PN, i.e., a nebula shining by reflected stellar light, or ionized by the passage of shocks and shining in forbidden lines. It is not clear why some post-AGB stars have a pre-PN and others do not.

\begin{figure}
\includegraphics{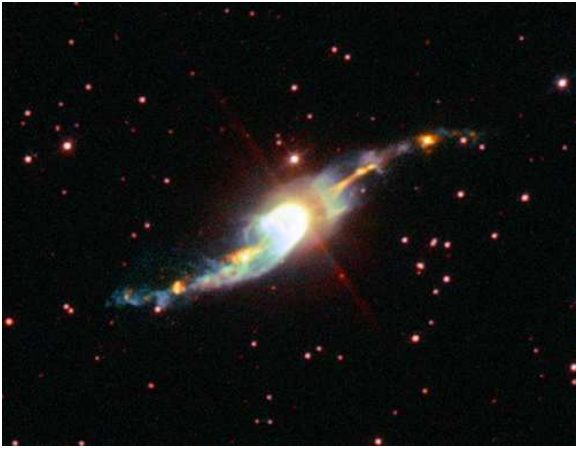}
\caption{The pre-PN He~3-1475, one of a sample studied by \citet{Huggins2007}. These objects are not yet photo-ionized, and are therefore thought to precede the PN phase; they always exhibit extreme morphologies. FOV$\sim$100''x80'', North is towards the top, East to the left (HST/WFPC2; image courtesy of  A. Riera and P. Garcia-Lario; yellow, orange, beige, green, blue: F814W,[NII],H$\alpha$,[SI],F555W).}
\label{fig:he3-1475}
\end{figure}

All post-AGB stars for which a nebula is observed display non-spherical morphologies \citep{Sahai2007}.  
In addition \citet{Huggins2007} carried out a kinematic analysis of 9 pre-PNe with tori and bipolar lobes (see Fig.~\ref{fig:he3-1475}), and showed that the lobes' ejection lags the torus ejection by $\sim$100 years. This observation is in line with a common envelope with a sub-stellar companion (see Sec.~\ref{ssec:primaryshapingagents}). In addition, \citet{Bujarrabal2001} studied a sample of pre-PNe and determined that $\sim$80\% of them have linear momenta (calculated from CO line measurements) much in excess (up to 1000 times) of what can be imparted by radiation pressure. This observation has been used to argue that a binary is needed to justify the linear momenta of these outflows.

{\it Post-AGB  binaries.}
Thirty percent of all post-AGB stars \citep[with no pre-PN;][]{vanWinckel2003} have thick circumstellar tori and almost every time one of these dusty post-AGB stars has been monitored spectroscopically, it has been found to be an intermediate period {\it spectroscopic} binary (100 $\la$ P $\la$ 1500 days, where for P$\ga$1~yr the ellipticities are larger than unity). These periods are too short for the companion to have remained outside the  AGB progenitor of the primary. Yet the period has clearly not been shortened dramatically by a common envelope interaction.  
If the post-AGB stars polled by \citet{vanWinckel2003} represent the CSPN progenitor population,  then one expects that 30\% of the CSPNe have companions at approximately similar separations (see \S~\ref{sec:predictionsandtests}). 

Searches for binaries in the middle of {\it pre}-PN have been carried out by \citet{Hrivnak2008}. Only pulsation-induced RV motion was detected in a small survey of 7 objects. The connection between the binary-rich post-AGB stars (with no pre-PN) and the apparent dearth of binaries in the pre-PN central star population is not clear. Clearly more data are needed for the pre-PNe. Some pre-PNe are known to contain binaries: the Red Rectangle has an A-type super-giant CSPN known to be a single-lined spectroscopic binary with a period of (319$\pm$5)~days \citep{Cohen2004}. The central system of the pre-PN OH~231.8+4.2 comprises a bright Mira \citep[QX~Pup;][]{Kastner1992} and an A star \citep{SanchezContreras2004}.  This object might however be a very peculiar system, possibly more related to symbiotic binaries (see also \S~\ref{ssec:symbiotics}). The pre-PN or very young PN GLMP 612 has a visual binary at its core \citep[][we counted this object among the visual binary CSPNe (Table~\ref{tab:widebinaries}), because there are arguments that the young nebula is photo-ionized, and is therefore an {\it bona fide} PN]{Garcia-Lario1997,Riera2003}.

\subsection{White dwarfs}
\label{ssec:wds}

Not all WDs have gone through an AGB phase. Some WDs derive from horizontal branch stars that did not have enough envelope mass to ascend the AGB. Furthermore, not all {\it post-AGB} WDs have gone through a PN phase, since those with M$\la$0.55~\msun\ are not massive enough to ionize the PN before the PN gas disperses. This is why \citet{Liebert2005} suggested that the WD birthrate density of 1.2$\times$10$^{-12}$ WD~pc$^{-3}$~yr$^{-1}$ is an absolute upper limit for the PN birthrate density. The latter number has been determined by several authors from local PN counts, and has been found to range between 0.8$\times$10$^{-12}$ WD~pc$^{-3}$~yr$^{-1}$ \citep{Frew2008} to a factor of a few larger than the WD birthrate density \citep[e.g.,][]{Phillips2002}, but it is seriously affected by the very uncertain distances to PNe.

{\it Binary WDs.}
From a volume-limited survey of 122 WDs, the WD binary fraction is 25\% \citep{Holberg2008,Holberg2008b}. This is thought to be a lower limit because many Sirius-type binaries, comprising a WD and a bright companion with spectral type earlier than G, would remain unidentified because the WD is outshone by the companion. About 8\% of all WDs are in Sirius-type binaries, so even an increase in the Sirius-like binary fraction is unlikely to bring the WD binary fraction much above the known 25\%. Considering the WD mass distribution and their mean-progenitor mass, one might expect a higher WD binary fraction, closer to that determined from main sequence stars of spectral type late-F to late-G, which is $\sim$60\% \citep{Duquennoy1991}. Even decreasing this number by a few percent to account for binaries that resulted in mergers, there still seem to be a deficit of WD binaries.  

\citet{Farihi2006} carried out an HST imaging survey of WDs known to have an IR excess from 2MASS colors. They found a bimodal distribution of projected separations that they interpreted as the common envelope period gap: binaries with a projected separation smaller than 13$^{+8}_{-3}$~AU (where the error bars are determined from the bin size in the histogram in Fig.~3 of \citet{Farihi2006}) enter a common envelope phase and have their periods decreased. \citet{Schreiber2008} and \citet{Schreiber2008b} determined that (35$\pm$11)\% of a sample of $\sim$100 WD+dM binaries from the SDSS are in post-common envelope systems (this translates to $\ga$9\% of all WDs are in post-CE systems, similar to population predictions -- see Sec.~\ref{ssec:fracanddist}). For all 28 systems measured, the period was found to be smaller than $\sim$1~day, where, had there been a sizable population of WD+dM binaries with periods between 1~day and 40 days, they would have detected them. This finding is in line with the dearth of post-common envelope CSPN binaries with 1-3~days $\la$P$\la$2 weeks (i.e., between the longest detected period and the reasonable upper period limit of the photometric survey - Sec.~\ref{ssec:binarycentralstarsearches}) and poses the question of whether the common envelope interaction might actually only produce {\it very} short period binaries \citep{DeMarco2009,Miszalski2009}. 

Finally, brown dwarf companions to WDs are very rare \citep{Farihi2005} and no planetary mass companion has ever been detected around a WD \citep{Farihi2008}, implying that those are rare too. However debris disks have been detected around some WDs \citep[e.g.,][]{vonHippel2007} implying that in some case remnants of possible planetary systems do survive stellar evolution. This information,  in combination with what is found for main sequence star, will play a role in the model of PN shaping.


\section{Binary central stars and related binary classes}
\label{sec:relatedclasses}

Here we list binary classes that are intimately related to CSPN binaries. The links can elucidate the evolution of these classes and mutually constrain theories for their formation (for an interesting review on the properties of evolved intermediate mass binaries see \citealt{Frankowski2007}).

\subsection{Symbiotic binaries}
\label{ssec:symbiotics}

\begin{figure}
\includegraphics{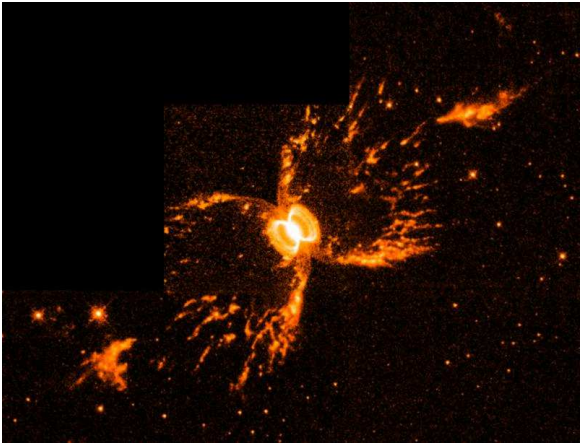}
\caption{The symbiotic nebula He~2-104, one of a sample studied by \citet{Corradi1993}. FOV$\sim$90''x60'', North is towards the top, East to the left (NTT/EMMI and HST/WFPC2; image courtesy of R. L. M. Corradi, M. Livio, U. Munari and H. Schwarz; [NII]).}
\label{fig:he2-104}
\end{figure}

Symbiotic binaries are intermediate period binaries (periods from one year to several decades; \citealt{Nussbaumer1996}), comprising a cool evolved star and an old WD kept hot by accretion of the cool star wind. In some cases they have resolved nebulae with ring-like or bipolar shapes (Fig.~\ref{fig:he2-104}). Symbiotic nebulae are different from PNe because they originate from the {\it cool star's} mass-loss and are ionized by the hot WD \citep[see, e.g.,][]{Podsiadlowski2007}, while bona fide PNe originate from the {\it hot star when it was on the AGB}, and are ionized by the hot star itself. In symbiotic nebulae we are witnessing the giant wind being shaped by the binary interaction. These systems therefore provide a wealth of information concerning mass-loss shaping. 

Symbiotic binaries might have an evolutionary connection to binary CSPNe: they could be their progenitors as well as progeny. The cool component in a symbiotic binary might make a PN in the future and the progenitor of the hot component is likely to have been a CSPN in the past. Symbiotic binaries and the post-AGB stars in binaries (Sec.~\ref{ssec:pagb}) argue for the fact that CSPNe with intermediate periods (one to several years) exist. 

The close relationship between symbiotic nebulae and PNe is a mixed blessing, since in order to use symbiotic systems to learn about PNe we need to be able to tell them apart. For instance, A~35-type central star binaries (Sec.~\ref{sec:theknownbinaries}) have an intermediate period, but the fact that the cool companion is evolved makes them suspect symbiotic systems themselves. The presence of an {\it evolved} cool companion in a CSPN binary should be regarded as suspicious, since the hot pre-WD primary was recently on the AGB. This implies that the two stars in the system had almost identical main sequence masses and equal mass-binaries should be rare \citep{Duquennoy1991}. This would automatically argue that such a binary inside a nebula is actually a symbiotic system. 

Of the 13 objects listed in Table~\ref{tab:coolcentralstars} with a known luminosity class, 6 are definitely evolved, indicating that they might be symbiotics. If one were to know the state of evolution of the hot star, this would provide us with a distinguishing characteristic: a PN should have a pre-WD or in any case hot WD central star, while a symbiotic nebula should have an older and cooler WD. However, the WD in a symbiotic system is kept warm by accretion of the giant's wind, and is therefore not trivial to establish its evolutionary stage. An alternative discriminant is the location of the giant on the HR-diagram. For instance, the cool central star of the PN Me~1-1, is estimated by \citet{Pereira2008}   to have just reached the base of the RGB; such star could not have lost enough mass to generate such a bright nebula; the nebula might therefore have been produced by the hot companion and be a PN. A PN interpretation for Me~1-1 is also supported by the fact that the cool star is displaced from the center of the nebula by 400~AU, which argues against being the origin of the nebula. 

Symbiotics also tend to have smaller nebular masses, although this is not always true \citep[e.g., the well-established D-type symbiotic nebula Hen~2-104 has a PN-like nebular mass close to 0.1~\msun;][]{Santander-Garcia2008}. \citet{Pereira2008} show a diagram of the line ratios [OIII] $\lambda$5007/H$_\beta$ and [OIII] $\lambda$4363/H$_\gamma$ where the loci of PN and symbiotics can be easily distinguished (Me~1-1 fall in this diagram at the bottom of the PN locus, although it is not distant from the leftmost side of the symbiotics locus).  \citet{Schmeja2003} argue that JHK color-color diagrams can distinguish symbiotic binaries from CSPN binaries, but this is also not foolproof (e.g., the central binary of PN NGC~2346 appears in the symbiotic region of such diagrams).

\subsection{Cataclysmic variables}
\label{ssec:CVs}

All post-common envelope CSPNe with a cool, main sequence companion can be considered pre-cataclysmic variables (CVs), since when the binary period decreases due to magnetic breaking and gravitational wave radiation \citep{Verbunt1981,Landau1962}, the system will enter contact, the main sequence star will transfer gas onto the WD and the system will commence CV activity. None of the post-common envelope systems in the middle of a PN exhibit CV activity \citep{DeMarco2008c}. However we know of two novae that took place in the middle of old PNe; Nova Vul 2007 No.~1 that went off in the middle of old PNe \citep{Wesson2008b} and nova GK~Per \citep{Bode1987}, which resides in an old nebula thought to be an old PN. However, pre-CV CSPNe provide a measure of the initial periods of systems that will eventually become CVs, and are therefore important when trying to understand period-altering mechanisms and the CV period distribution.


Finally, we would like to draw attention to another connection between PNe and CVs. {\it Born-again} CSPNe are commonly attributed to a final helium shell flash in a single post-AGB star \citep[][see also \S~\ref{ssec:hyddef}]{Herwig2001}. The ejecta from the outburst are expected to be carbon-rich. \citet{Wesson2003} and \citet{Wesson2008} analyzed the ejecta abundances of two (A~30 and A~58) of the five known born-again CSPNe  and found both to have oxygen and neon-rich compositions. This is in glaring contradiction with predictions from the classical born-again scenario and in line with the abundances of oxygen-neon-mgnesium novae, establishing a puzzling connection between this rare class of novae and the born-again phenomenon. Though no scheme can currently explain this connection, nor other puzzling characteristics of these stars, an alternative to the simple born-again scenario, involving close binaries is presented by \citet{DeMarco2008b}. 

\subsection{Binary post-RGB stars}
\label{ssec:postRGBs}

Hot horizontal branch stars (sdB stars) have been shown to be, by and large, binary systems that went through a common envelope phase on the RGB \citep{Maxted2001,MoralesRueda2004}. 
The sdB binary period distribution peaks at less than 1~day, but has a substantial number of systems with periods as long as $\sim$100~days, contrary to what is observed in the post-common envelope CSPNe and WD populations (Sections~\ref{ssec:binarycentralstarsearches} and ~\ref{ssec:wds}). This discrepancy might reveal important details of the common envelope interaction.  

\subsection{Other post-AGB binary-related classes}
\label{ssec:otherclasses}
\citet{Frankowski2007} give a list of single stars and binaries that are related to binary AGB and post-AGB stars (such as barium or CH stars). Some of these stars are thought to have merged with a companion. They propose a scheme to accommodate all of these classes. Testing such a scheme would provide a framework to relate all these stellar classes and to make the best use of the stellar characteristics of each class in the understanding of another.



\section{Additional riddles facing any successful model for PN origin and shaping}
\label{sec:additionalriddles}

Before concluding, it is opportune to list a few more issues facing the  Single Star Paradigm. The Binary Hypothesis or any other successful model has to provide an answer to the riddles below.  

\subsection{The PN luminosity function (PNLF)} 

There is no physical reason why the bright end of the PNLF should be a standard candle: in older galaxies the bright edge of the PNLF should be 4 magnitudes fainter than in younger ones \citep{Marigo2004}. Yet the PNLF is extremely successful in predicting distances to old and young galaxies alike \citep{Jacoby1992}. \citet{Ciardullo2005} suggested that the bright edge of the PNLF in older populations is populated by stars that have a higher mass because they suffered a merger. This theory is similar to that proposed to explain blue strugglers in clusters. An alternative model to explain the PNLF involving symbiotic binaries is put forward by \citet{Soker2006b}.


\subsection{Hydrogen-deficient central stars}
\label{ssec:hyddef}  

About 10-20\% of all CSPNe are hydrogen-deficient, the product, it is believed, of a final thermal pulse that took place when the CSPN was already on the WD cooling track, and that lead to a re-expansion and a new, hydrogen-deficient PN \citep[][these stars are also known as born-again stars]{Herwig2001}. Most of these stars exhibit emission line spectra that mimic those of population I Wolf-Rayet stars \citep{Crowther1998}. Today we have identified five objects believed to have gone through a final thermal pulse (three, Sakurai's Obsect, V~605~Aql and FG~Sge, were caught during the outburst, while the other two, A~30 and A~78, are CSPNe known to be hydrogen deficient {\it and} that have hydrogen-deficient circumstellar ejecta; for a review, see, e.g., \citealt[][]{Clayton1997}).  

However, this scenario cannot explain all the characteristics of the hydrogen-deficient CSPN population. For instance, (i) there is a suspiciously high abundance of the usually rare hydrogen-deficient PNe in environments with {\it few} PN overall (such as GCs and the Sagittarius Dwarf Galaxy; \citealt{Zijlstra2001}). (ii) The hydrogen-deficient ejecta of A~30 and A~58 are oxygen- instead of carbon-rich \citep{Wesson2003,Wesson2008}, as instead predicted by the final thermal pulse theory (\S~\ref{ssec:CVs}). (iii) The evolutionary sequence of Wolf-Rayet CSPNe, while well established from the stellar abundance point of view \citep[e.g.,][]{Crowther1998}, is not so obvious when we look at other characteristics, such as the number of stars in each Wolf-Rayet subclass \citep[e.g.,][]{Zijlstra2001}. \citet{DeMarco2002} and \citet{DeMarco2008b} suggests scenarios involving mergers or a final thermal pulse in a close binary, but it has to be admitted that no scenario currently succeeds in explaining all the observations.  

\subsection{Faint circular PNe} 

An increasing number of perfectly circular PNe (e.g., Fig.~\ref{fig:abell39}) are being found by surveys that detect very faint PNe \citep[e.g.,][]{Pierce2004,Parker2006,Miszalski2008b}. The riddle in this discovery is that all these circular PNe are relatively old. Where are their young counterparts? This new PN population lead George Jacoby to suggest, during a discussion at IAU Symposium 234 \citep{Barlow2006} that PNe are a mixed class, whose members have evolved through a variety of evolutionary channels. 


\section{Conclusions}
\label{sec:conclusions}

The appeal of the Binary Hypothesis on theoretical grounds is undeniable. Population synthesis shows that a large fraction of the PN population deriving from binary interactions is {\it consistent} with what we know of binarity, stellar evolution and galactic history. However, so far neither the Single Star Paradigm, nor the Binary Hypotheses are clearly supported by evidence.  

Detecting the binary fraction and period distribution of CSPNe for the {\it entire period range} has to take the utmost priority. From photometric variability surveys we have a good idea that the post-common envelope binary fraction is only 12-21\%, and that the common envelope evolutionary channel does indeed lead to very short period binaries, with very few systems in the period range 3 days-2 weeks (where 2 weeks is a reasonable upper limit for this survey technique). 

We still do not know whether there are any binaries with periods longer than can be probed with the photometric technique. RV surveys, as we have seen, are fraught with problems, since variable winds can induce RV variability. It is possible that observations of faint objects likely to have very weak winds might lead to some detections. 
The near-IR excess method is itself problematic, since hot CSPNe are bright even at near-IR wavelengths, but still might be the best way to find the elusive intermediate period binaries.  
For binaries with separations between about one and 50-100~AU it is possible that adaptive optics and interferometry will start finding companions in the near future, as these techniques are quickly being improved \citep[e.g., $\delta$~Cen][]{Melliland2008}.

Deep, time-dependent, large scale surveys such as Pan-STARRS\footnote{pan-starrs.ifa.hawaii.edu.}, Sky Mapper\footnote{msowww.anu.edu.au\/skymapper.} or the Large Synoptic Survey Telescope\footnote{www.lsst.org.} are coming on line in the next decade and will be able to quantify variability in a much larger number of CSPNe. This alone will dramatically increase the size sample of irradiated binaries so that they can be treated statistically.  

It is possible that mergers and sub-stellar companions play a much larger role in shaping PNe than we currently assume. This issue  will have to be resolved indirectly, by associating observable characteristics to the presumed merging events that took place (for instance \citet{Taut2008} found strong evidence that single magnetic WDs derive from mergers) or to the presumed interaction with a planetary companion. While this might take us back into the land of circumstantial evidence, the key will be obtaining large numbers of objects with characteristic ascribed to a given merger event or planetary interaction type. Population statistics will then come to bear. Once again, large surveys will multiply the number of peculiar systems and knowledge of brown dwarf and planetary companions around main sequence and evolved stars will fill in the blanks in the overall picture of the role of companions in stellar evolution and in the shaping of PNe.

PN mimics might constitute a much larger fraction of the current PN population than previously thought \citep{Frew2008b,Miszalski2009}. We consider mimics nebulae which do not derive from the atmosphere of the hot star causing the ionization. These must be identified before we can understand PN phenomenon. 

Finally, we should keep in mind that even in the Single Star Paradigm one out of $\sim$5 PNe has been generated by a common envelope interaction, a very different PN ejection mechanism to the single star ejection that is presumed to generate the other PNe. This fact alone shows us that PNe derive from at least two distinct evolutionary channels. What is left to be determined is just {\it how many} other evolutionary channels there are, and which ones involve a companion.

 \acknowledgments
For critical comments on drafts of this paper or insightful discussions I acknowledge Noam Soker, David Frew, Jason Nordhaus, Bruce Balick and Brent Miszalski. I am also grateful for helpful comments to the manuscript by Mike Barlow, Geoff Clayton, George Jacoby, Thomas Rauch, Raghvendra Sahai, Albert Zijlstra and the referee, Julie Lutz. For insightful conversations on WDs I thank Jay Farihi, Matthias Schreiber, Alberto Rebassa-Mansergas and Jay Holberg. For advice on planets and brown dwarfs around main sequence stars I am grateful to Jackie Faherty and Ben Oppenheimer. I am finally grateful for the well timed input of Joel Kastner who, at the Asymmetric PN IV meeting, quoting the late Hugo Schwarz's words spoken at a previous meeting, encouraged the community to coordinate and find an observational test for the Binary Hypothesis. His speech marked the start of PlaN-\"B and of this effort. This work has been supported by NSF grant AST-0607111.

 \bibliographystyle{../../apj}                       
\bibliography{../../bibliography}






\clearpage

\begin{table*}
\caption{RV surveys of central stars of PNe. The brightness limit of these surveys is $\sim$14-15~mag.}
\begin{tabular}{lcccccc}
\hline
RV survey         & Res.  &Telescope &	\# 	&\#          & SNR$^1$&	\% RV  	  \\
                         & (\AA) &                 &  Obj.   &  Meas. &        &Variables \\
\hline
\citealt{Mendez1989}	& 0.3	          & 3.6-m ESO  	&28	& 1-2   & 	--        &      0\%          \\
\citealt{Sorensen2003}	& 1.5                   & 2.5-m INT	         & 33	& 6-40 & ~100   &	39\%        \\  
\citealt{DeMarco2004}	& 0.6  & 3.6-m WIYN 	& 11 & 6-16 & ~30-50&	91\%         \\
 \citealt{Afsar2005}	& 1.5  & 1.5-m CTIO        & 19 &  5-47 & ~30-50& 	37-50\%  \\
  \hline
\multicolumn{7}{l}{$^1$Signal-to-noise ratio.}
\end{tabular}
\label{tab:RVsurveys}
\end{table*}

\begin{table*}\def~{\hphantom{0}}
  \begin{center}\begin{tiny}
  \caption{Known ``closer" binary central stars (references in \citealt{DeMarco2008c}, Sec. \ref{sec:theknownbinaries} and in \citet{Miszalski2009}). Visual binaries are listed in Table \ref{tab:widebinaries}.}
  \label{tab:knownbinaries}
  \begin{tabular}{lllllllllll}\hline
      PN name     &Type$^1$ &Period&   $M_1^2$ & $M_2^2$ & $T_1$   & $T_2$  &$R_2$& $i^2$ & Spec. Type$^3$ &Morphology$^4$  \\
                         &&(days)&  (\msun)   &(\msun)       & (kK) & (K)  &(\rsun)& (deg) & Secondary    &    \\
                         \hline
Pe~1-9$^5$            & Ec,El? & 0.14 &  -- & -- & -- & -- & -- & -- & -- &?\\
BMP~1800-3408$^5$& Ec,I & 0.14 &  -- & -- & -- & -- & -- & -- & -- &--\\
PNG136$^{6}$     & S1,El & 0.16 &0.55&$>$0.82&120\,000&--&--&--&WD/NS&J\\ 
	      	             &         & 0.23 &{\it 0.6}&{\it 0.30}&50\,000$\pm$5000&46\,000$\pm$4000&0.58&67$\pm$2&sdB &\\
NGC~6337$^{6}$	     & I       & 0.17 &{\it 0.6}& 0.35 &45\,000  & 5500 &0.42&28 &M5V&R:J:\\ 
                            &         &        &{\it 0.6}& 0.20 &105\,000& 2300 &0.34&9 &M3V& \\                                                 
PPA~1747-3435$^5$    & I & 0.22 &  -- & -- & -- & -- & -- & -- & -- &--\\
A~41	             & El,S1& 0.23 &0.56$\pm$0.08&0.56$\pm$0.08&45\,000$\pm$5000&40\,000$\pm$4000&0.71$\pm$0.08&66$\pm$2&sdB&W:\\ 
H~2-29$^5$                 & Ec,El? & 0.24 & -- & -- & -- & -- & -- & $\sim$90 & -- &W\\
PHR~1756-3342$^5$    & I & 0.26 &  -- & -- & -- & -- & -- & -- & -- &--\\
Bl~3-15$^5$                & El & 0.27 &  -- & -- & -- & -- & -- & -- & -- &--\\
JaSt~66$^5$                & El & 0.28 &  -- & -- & -- & -- & -- & -- & -- &--\\
Sab~41$^5$                  & I & 0.30 &  -- & -- & -- & -- & -- & -- & -- &--\\
PPA~1759-2834$^5$    & I & 0.31 &  -- & -- & -- & -- & -- & -- & -- &--\\
PHR~1801-2947$^5$     & I & 0.32 &  -- & -- & -- & -- & -- & -- & -- &--\\
PHR~1801-2718$^5$     & I & 0.32 &  -- & -- & -- & -- & -- & -- & -- &--\\
DS~1	             & S2,I  & 0.36 &{\it 0.63$\pm$0.03}&0.23$\pm$0.01&77\,000$\pm$8000&3400$\pm$1000&0.402$\pm$0.005&62.5$\pm$1.5&M5V&?\\ 
K~6-34$^5$                  & El & 0.39 &  -- & -- & -- & -- & -- & -- & -- &--\\
Hf~2-2		     & I?      & 0.40&--&--&$\sim$67\,000&--&--&--&--&R:\\
A~63$^7$	             & Ec,I   & 0.46&0.63$\pm$0.06&0.29$\pm$0.04&$\sim$117\,500$\pm$12\,500&7300$\pm$250&0.53$\pm$0.02&88&M4V&WJ\\ 
A~46	             & S2,Ec$^7$,I&0.47&0.51$\pm$0.07 & 0.15$\pm$0.02 &60\,000$\pm$10\,000 & 5300$\pm$500 &0.46$\pm$0.03 & 80.5$\pm$0.2 &M6V&? \\ 
MPA~1759-3007$^5$    & El & 0.50 & -- & -- & -- & -- & -- & -- & -- &--\\
M~3-16$^5$                & Ec,El & 0.57 & -- & -- & -- & -- & -- & $\sim$90 & -- &RJ\\
HFG~1$^6$		  &S2,I,El&0.58&{\it 0.57}&1.09&83\,000$\pm$6000 &-- &--&28$\pm$2 & F9V &J\\ 
     	           &	&  &{\it 0.63}&0.41&83\,000$\pm$6000 &-- &--&29 & M2V&\\ 
NGC~6026         &El&0.58 & 0.53$\pm$0.01&0.53$\pm$0.01&36\,000&134\,000$\pm$5000&0.053$\pm$0.005&82$\pm$5&WD/sdO&R::\\
PHR~1804-2645$^5$  & Ec & 0.62 &  -- & -- & -- & -- & -- & -- & -- &--\\
M~2-19$^5$              & El & 0.67 & -- & -- & -- & -- & -- & $\sim$90 & -- &RB\\
K~1-2	          &S,I&0.68&{\it 0.6}&$>$0.74&$\sim$85\,000&--&--&{\it 50}&earlier than K2V&J\\ 
PHR~1757-2824$^5$ & Ec,I & 0.80 &  -- & -- & -- & -- & -- & -- & -- &--\\
A~65	          &I&1.00$^9$&--&--&$\sim$80\,000&--&--&--&--&R::W::\\
PHR~1759-2915$^5$  & Ec & 1.10 &  -- & -- & -- & -- & -- & -- & -- &--\\
H~1-33$^5$               &I& 1.13 &  -- & -- & -- & -- & -- & -- & -- &--\\
HaTr~4		  &I&1.74&--&--&--&--&--&--&--&W\\                          
BE~UMa$^{10}$     &S2,Ec,I&2.29&0.70$\pm$0.07&0.36$\pm$0.07&105\,000$\pm$5000&5800$\pm$300&0.72$\pm$0.06&84$\pm$1&M3V&W\\  
Sp~1    		  &I&2.91&--&--&--&--&--&--&--&R\\ 
SuWt~2$^{11}$		&S2&4.9&$\sim$2.5&$\sim$2.5&--&--&--&$\sim$90&A & RB\\
PHR 1804-2913$^5$       & I & 6.66 &  -- & -- & -- & -- & -- & -- & -- &--\\
PHR~1744-3355$^5$     &I & 8.23 & -- & -- & -- & -- & -- & -- & -- &--\\
NGC~2346	&S1&15.99&0.40$\pm$0.05&1.8$\pm$0.3&--&8000&--&$>$50&A5V&B\\
   A~35$^{12}$ &C&--&$\sim$0.5&--&80\,000$\pm$3\,000&5200$\pm$200&--&--&G8~III-IV&--\\
  LoTr~1 &C&--&--&--&v. hot&--&--&--&K&--\\
  LoTr~5 &C&--&--&--&150\,000&5230&--&--&G5III&--\\
  NGC~1514 &C&--&--&--&$>$60\,000&--&--&--&A0-3~III&--\\
  NGC~2438 &C&--&0.56$\pm$0.01&--&114\,000$\pm$10\,000&3470&--&--&M3V&--\\
    \hline
   \multicolumn{11}{l}{$^1$Legend: S1, S2: single or double-lined spectroscopic binaries; Ec: eclipsing; El: ellipsoidal variability; I: irradiated; C: composite spectrum or colors.}\\
   \multicolumn{11}{l}{$^2$Values in italics are used for assumed quantities.}\\
   \multicolumn{11}{l}{$^3$For irradiated and eclipsing systems this is the spectral type corresponding to a main sequence star of the determined mass.}\\
   \multicolumn{11}{l}{$^4$Legend: B: clear, bipolar lobes; R: clear ring; W: very likely that PN is the edge-on waist of a faded bipolar; J: presence of one or a pair of jets or jet-like structures.}\\
\multicolumn{11}{l}{\ \  A ``:" means that the attribution is uncertain, a ``::" means it is very uncertain, a ``?" means that data exist but the morphology is hard to classify.}\\  
   \multicolumn{11}{l}{$^5$These are binary systems newly discovered by the survey of \citet{Miszalski2009}.}\\
     \multicolumn{11}{l}{$^{6}$Two models can reproduce the data equally well; see discussion in \citealt{DeMarco2008c}.}\\
    \multicolumn{11}{l}{$^7$This close binary central star might have a wide companion as well \citep[][and Table \ref{tab:widebinaries}]{Ciardullo1999}.}\\
  \multicolumn{11}{l}{$^8$Partly eclipsing.} \\
  \multicolumn{11}{l}{$^9$The period estimate was recently refined by D. Frew and T. Hillwig, priv. comm.}\\
    \multicolumn{11}{l}{$^{10}$BE~UMa is the name of the central star, the PN is PN G144.8+65.8}.\\
   \multicolumn{11}{l}{$^{11}$This central star is a double-lined spectroscopic binary where both stars are of type A!}\\
   \multicolumn{11}{l}{$^{12}$\citet{Frew2008b} disputes the {\it bona fide} origin of this PN and argues instead that it is a Str\"omgren sphere around a binary, whose hot component was a central star}\\ 
   \multicolumn{11}{l}{\ \ \ in the recent past. The binary is resolved (separation 0.08-0.14\arcsec), but the period is not known.}\\
   \end{tabular}
 \end{tiny} \end{center}
\end{table*}

\clearpage

\begin{table*}\def~{\hphantom{0}}
  \begin{center}
  \caption{Known visual binary central stars, mostly from \citealt{Ciardullo1999}. For other references see Sec. \ref{sssec:widebinaries}.}
  \label{tab:widebinaries}
  \begin{tabular}{lllll}
  \hline
      PN name     &Association   & Sep.&   Comments  \& companion & Ref.\\
                         &   is$^1$:             &         (arcsec)  &   spectral type$^2$ &      \\
                         \hline
GLMP 621  &--&1& possibly a pre-PN& \citealt{Riera2003}\\
 EGB~6 &--&0.18& (M5V)&\citealt{Bond1993}\\
 NGC~246&--&3.8& G8-K0V (K0V)&\citealt{Bond1999}\\
  He~3-1357 &--&0.4& late spectral type &\citealt{Bobrowsky1998} \\  
  NGC~6818 & --& 0.09 &G8-K0 &\citealt{Benetti2003}\\
  A~31 &``Probable"&0.26& later than M4 (M6V)& \citealt{Ciardullo1999}\\
  A~33 &``Probable"&1.82& K2 (K3V)& \citealt{Ciardullo1999}\\
  K~1-14 &``Probable"&0.36& K0& \citealt{Ciardullo1999}\\
  K~1-22 &``Probable"&0.35& K1 (K1V)& \citealt{Ciardullo1999}\\
  K~1-27 &``Probable"&0.56& likely WD companion& \citealt{Ciardullo1999}\\
  Mz~2 &``Probable"&0.28& K2& \citealt{Ciardullo1999}\\
  NGC~1535 &``Probable"&1.04& G8& \citealt{Ciardullo1999}\\
  NGC~3132 &``Probable"&1.71& A3 (A2 IV-V)& \citealt{Ciardullo1999}\\
  NGC~7008 &``Probable"&0.42& K3 companion is close binary (GV:)& \citealt{Ciardullo1999}\\
  Sp~3         &``Probable"&0.31& F9& \citealt{Ciardullo1999}\\
  A~7 &``Possible" & 0.91& later than K2 (M4V)& \citealt{Ciardullo1999}\\
 A~30 &``Possible" & 5.25 & G8& \citealt{Ciardullo1999}\\
 A~63 & ``Possible"& 2.82& K1& \citealt{Ciardullo1999}\\
 IC~4637 &``Possible"& 2.42& K4& \citealt{Ciardullo1999}\\
 NGC~2392 &``Possible" & 2.65& later than F2 (MV::)& \citealt{Ciardullo1999}\\
 NGC~2610 &``Possible" & 0.61& later than G0& \citealt{Ciardullo1999}\\
 \hline
   \multicolumn{5}{l}{$^1$``Probable" and ``Possible" are the adjective used by  \citet{Ciardullo1999} to describe the}\\ 
    \multicolumn{5}{l}{  association probability of these systems. ``Doubtful" systems have not been included as there are}\\ 
    \multicolumn{5}{l}{  several arguments against the assocaition.}\\ 
     \multicolumn{5}{l}{$^2$ Here we report the spectral type of the companion from the literature (see \S~\ref{sssec:widebinaries}).}\\
      \multicolumn{5}{l}{ Characters in parenthesis are the spectral types determined by \citet{Frew2008b}.}\\
   \end{tabular}
 \end{center}
\end{table*}

 \clearpage
  
  \begin{table}\def~{\hphantom{0}}
  \begin{center}\begin{tiny}
  \caption{PNe with ``cool" central stars.}
  \label{tab:coolcentralstars}
  \begin{tabular}{llllllll}\hline
      PN name     &Spectral  & PN            &D$_{\rm PN}$&$N_e$& $T_{WD}$   &Comments &Reference    \\
                         &  Type    &  Morphology &(arcsec)&(cm$^{-3}$)&    (kK)  &&   \\
                         \hline
A~14 & B5 III-V & elip./bip. &33&$<$100&-- &born-again?& \citealt{Abell1966,Lutz1987}\\
A~79 &F0 V&bipolar&54&100&165& --&\citealt{Rodriguez2001,Bohigas2008}\\
A~82& K0 IV & elliptical& 55& 50&-- &--&\citealt{Ciardullo1999}\\
H~3-75   & G-K & round & 24 &-- & --& no UV excess; born-again? & \citealt{Bond2002b}\\
He~2-36 &A2 III&bip./ell.&40$\times$20&700&50--70:  &very likely binary& \citealt{Feibelman2001,Lee2007}\\
He~3-1312 & F(6-7) I& bipolar &16&--&--&post-AGB? symbiotic?&\citealt{Pereira2004}\\ 
Hu~1-1 & cool &  elliptical &5&1320&100& -- &\citealt{Kaler1976,Manchado1996}\\
&&&&&&&\citealt{Sterling2008}\\
IC~2120   & G & round & 42 & --&-- & no UV excess; born-again? & \citealt{Bond2002b}\\
               &&&&&&HII region?&\citealt{Zijlstra1990}\\
IRAS~19127$^1$ &B9V&stellar&--&$\sim 10^6$& --&symbiotic?&\citealt{Whitelock1986,Frew2008b}\\
&&&&&&&\citealt{Gauba2003}\\
K~1-22 & F V & elliptical & 200 & 21 & hot  &--& \citealt{Rauch1999}\\
K~3-43     & M &  -- &3&-- &--&poss. misclass.& \citealt{Sabbadin1987} \\
K~3-27     & G0: & bipolar& 8 &-- &hot &-- & \citealt{Perek1967}\\
               &&&&&&&\citealt{Manchado1996 ,Phillips2003}\\
Me~1-1 & K(1-2) II & bipolar&8$\times$2&$10^4$&62 &symbiotic?& \citealt{Shen2004, Pereira2008} \\
NGC~2346$^2$&A5 V&bipolar&52&500&100&-- &\citealt{Mendez1981, Su2004}\\ 
NGC~2899 & F  V:  &bipolar &140x69&126&$>$110:&bent lobes&\citealt{Gorny1999,Rauch1999}\\
Sa~3-151 & A &stellar?&--&--&--&little information&Acker, priv. comm.\\
SaSt~1-1 &G2 III/G5&elliptical&10&600&80 &D' symbiotic? barium star? &Schwarz 1991; \citealt{Pereira2005}\\
SuWt~2$^2$ &A+A&ring&41&100&--&-- &\citealt{Smith2007,Bond2002}\\
VBRC~1 & F V: & bipolar & 68 & 49 & -- & -- &\citealt{Rauch1999}\\
WeBo~1 & K0 III& ring &25&312&--&barium star&\citealt{Bond2003, Smith2007}\\

\hline
\multicolumn{8}{l}{$^1$IRAS~19127--1717}\\
\multicolumn{8}{l}{$^2$NGC~2346 has a single-lined spectroscopic binary CSPN of spectral type A, while SuWt~2 has a double-lined spectroscopic binary CSPN where}\\ 
\multicolumn{8}{l}{both components have spectral type A -- a third hot star is therefore suspected (Table~\ref{tab:knownbinaries}).}\\
  \end{tabular}
\end{tiny} \end{center}
\end{table}

\vspace{9cm}

\vspace{8cm}

\clearpage
\vspace{5cm}

\vspace{10cm}

\clearpage 

\vspace{10cm}

\clearpage
\vspace{12cm}


\clearpage
\vspace{6cm}

\clearpage
\vspace{5cm}

\vspace{5cm}

\end{document}